# Low rank spatial econometric models

Daisuke Murakami[1,*], Hajime Seya[2], Daniel A. Griffith[3]

[1] Department of Data Science, Institute of Statistical Mathematics,

10-3 Midori-cho, Tachikawa, Tokyo, 190-8562, Japan (Email: dmuraka@ism.ac.jp)

[2] Department of Civil Engineering, Graduate School of Engineering, Kobe University,

1-1 Rokkodai, Nada, Kobe, 657-8501, Japan (Email: hseya@people.kobe-u.ac.jp)

[3] School of Economic, Political and Policy Science, The University of Texas at Dallas,

800 W Campbell Rd, Richardson, TX, 75080, USA (Email: dagriffith@utdallas.edu)

[*]Corresponding author

**Abstract**: This article presents a re-structuring of spatial econometric models in a linear mixed model framework. To that end, it proposes low rank spatial econometric models that are robust to the existence of noise (i.e., measurement error), and can enjoy fast parameter estimation and inference by Type II restricted likelihood maximization (empirical Bayes) techniques. The small sample properties of the proposed low rank spatial econometric models are examined using Monte Carlo simulation experiments; the results of these experiments confirm that direct effects and indirect effects à la LeSage and Pace (2009) can be estimated with a high degree of accuracy. Also, when data are noisy, estimators for coefficients in the proposed models have lower root mean squared errors compared to conventional specifications, despite them being low rank approximations. The proposed approach is implemented in an R package "spmoran."

# 1. Introduction

Both spatial econometrics (e.g., LeSage and Pace, 2009; Kelejian and Piras, 2017) and geostatistics/spatial statistics (e.g., Cressie, 1993; Cressie and Wikle, 2011) offer useful toolboxes for spatial and spatiotemporal data modeling. The model specifications characterizing these two fields are quite different. Recent studies in geostatistics, including Cressie and Wikle (2011), advocate the following three-stage hierarchical specification of Berlinear (1996):

- Data model        : [ data | spatial process, data parameters]

- Process model     : [ spatial process | process parameters]

- Parameter model   : [ data parameters, process parameters]

where the data model describes a data distribution conditional on the underlying process, the process model describes an underlying spatially dependent process, and the parameter model adds structures to a set of parameters. Geostatistical studies have shown the flexibility of this specification (e.g., Kang and Cressie, 2011; Banerjee et al., 2014).

In contrast, spatial econometric models typically assume that a spatial autocorrelation effect is present in a researcher's data sample, rather than in the population from which the sample was drawn (see Manski, 1993). In other words, associations among observations (i.e., "data") are modelled[1], and the distinction of *Data model* and *Process model* is not obvious compared to a geostatistical specification. For example, Arbia (2006) notes that the spatial lag model (SLM) (or spatial autoregressive response model), one of the popular model specifications in spatial econometrics, in which spatial autocorrelation within a response variable is modeled, is not consistent with random field theory.

Our aim in this article is to re-structure spatial econometric models in a linear mixed model framework, and explicitly distinguish the *Data model* and the *Process model*. By doing so, we try to make spatial econometric models more compatible with those used by statisticians working in geostatistics and/or other spatial statistical areas. Modelling within a linear mixed model framework allows natural dimension reduction of spatial econometric models. To that end, we propose low rank spatial econometric models that are robust to the existence of noise (i.e., measurement error), and can enjoy fast parameter estimation and inference by Type II restricted likelihood maximization (empirical Bayes) techniques. In geostatistics, low rank modeling is increasingly popular because of its computational efficiency, flexibility, and affinity with the hierarchical representation. Typical low rank approaches include fixed rank kriging (Cressie and Johanesson, 2008), predictive process modeling (Banerjee et al., 2008), and multiscale approximation (Nychka et al., 2015). However, approaches for low rank approximation of spatial econometric models are quite limited. One notable exception is Burden et al. (2015), who derived a low rank specification for the spatial error model (SEM) (also known as the simultaneous autoregressive model, in the spatial statistical literature). The difference between Burden et al. (2015) and our approach is that the latter is based on eigenvector spatial filtering (Griffith, 2003; Griffith and Chun, 2014), and it proposes not only a low rank SEM (LSEM), but also a low rank SLM (LSLM). In addition, we show that marginal effects, known as "direct effects" and "indirect effects" (LeSage and Pace, 2009), can naturally be defined and accurately calculated. Then, the small sample properties of the proposed low rank spatial econometric models are examined using Monte Carlo simulation experiments; the results of these experiments confirm that when data are noisy, coefficient estimators in the proposed models have lower root mean squared errors (RMSE) compared to conventional specifications, despite their being low rank approximations.

---

[1] This specification has merit in terms of identification of different types of spatial effects (see Gibbons and Overman, 2012).

The subsequent sections of this paper are organized as follows. Section 2 introduces the standard spatial econometric modeling approach. Section 3 develops the low rank spatial econometric model, accompanied by an explanation of its parameter estimation procedure. Section 4 compares the performance of conventional and our proposed low rank mixed model specifications using Monte Carlo simulation experiments. Finally, Section 5 concludes our paper.

## 2. Standard spatial econometric models

Spatial econometric models are autoregressive in nature, tend to be based upon the auto-normal specification, frequently have their parameters estimated with maximum likelihood techniques, and involve matrix inversions that generate direct and indirect effects. Meanwhile, the issue of dimensionality refers to the number of observations, $N$. As $N$ increases, the computational cost for the inversions increases rapidly. Low rank approach reduces the cost by approximating the (full rank) matrix being inverted by a rank $L$ ($< N$), or low rank, matrix, which allows for estimating parameters computationally efficiently. This section introduces standard spatial econometric models, and the subsequent section applies a lank reduction to these models.

### 2.1. Model specification

The SLM and the SEM are basic spatial econometric models formulated as follows (see Cliff and Ord, 1981; Upton and Fingleton, 1985; Anselin, 1988):

$$\mathbf{y} = \mathbf{X}\boldsymbol{\beta} + \mathbf{e} \quad \mathbf{e} = \varphi\mathbf{W}\mathbf{e} + \boldsymbol{\varepsilon} \quad \boldsymbol{\varepsilon} \sim N(\mathbf{0}, \sigma^2\mathbf{I}), \tag{1}$$

$$\mathbf{y} = \rho\mathbf{W}\mathbf{y} + \mathbf{X}\boldsymbol{\beta} + \boldsymbol{\varepsilon} \quad \boldsymbol{\varepsilon} \sim N(\mathbf{0}, \sigma^2\mathbf{I}), \tag{2}$$

where $\mathbf{y}$ is a vector of response variables ($N \times 1$), $\mathbf{X}$ is a matrix of covariates ($N \times K$), $\boldsymbol{\beta}$ is a vector of regression coefficients ($K \times 1$), $\sigma^2$ is a variance parameter, and $\mathbf{I}$ is an $N \times N$ identity matrix. $\mathbf{W}$ is an $N \times N$ matrix describing spatial connectivity among sites on which sample values locate. It is usually given by a row-standardized matrix calculated from a symmetric proximity matrix $\mathbf{W}_0$ with zero diagonals. Alternatively, $\mathbf{W}$ can be given by a symmetric matrix (see Elhorst, 2014), which is the version we will employ later. For example, $\mathbf{W}_0$ may be given by a rook-type proximity matrix whose ($i, j$)-th element equals 1 if these two samples are adjacent, and 0 otherwise. $\varphi$ and $\rho$ are parameters quantifying the strength of spatial dependence in **y** or **e**, respectively. Positive spatial dependence exists if these parameters are positive and significant, whereas negative dependence exists if they are negative and significant. The spatial dependence parameters take a value in the interval $(1/\lambda_{min}, 1/\lambda_{max}) = (1/\lambda_{min}, 1)$, where $\lambda_{max}$ and $\lambda_{min}$ are the maximum and the minimum eigenvalues of matrix **W**.

The SAC model and the spatial Durbin model (SDM) are other popular specifications in spatial econometrics. The SAC model, which combines the SEM and the SLM, is formulated as

$$\mathbf{y} = \rho\mathbf{Wy} + \mathbf{X}\boldsymbol{\beta} + \mathbf{e} \quad \mathbf{e} = \varphi\mathbf{We} + \boldsymbol{\varepsilon} \quad \boldsymbol{\varepsilon} \sim N(\mathbf{0}, \sigma^2\mathbf{I}). \tag{3}$$

This model is useful to estimate direct and indirect effects in the presence of residual spatial dependence (see Elhorst, 2014). The SDM is formulated as

$$\mathbf{y} = \rho\mathbf{Wy} + \mathbf{X}\boldsymbol{\beta} + \mathbf{WXq} + \mathbf{e} \quad \boldsymbol{\varepsilon} \sim N(\mathbf{0}, \sigma^2\mathbf{I}), \tag{4}$$

where **q** is a vector of coefficients. The **WXq** term captures local spill-overs from **X** while the $\rho\mathbf{Wy}$ term captures global spill-over effects (see Elhorst, 2009; Seya et al., 2012).

These spatial econometric models are widely used, especially in the social sciences. See Anselin (2010) about the history of spatial econometrics.

## 2.2. Maximum likelihood (ML) estimation

Spatial dependence parameters in Eqs. (1) – (4) are estimated by, for example, maximizing their log-likelihood functions, which are specified as follows:

$$\log p(\mathbf{y}|\boldsymbol{\theta}) = const. - \frac{N}{2} \log\left[\frac{1}{N}\hat{\mathbf{r}}'\mathbf{Z}'\mathbf{Z}\hat{\mathbf{r}}\right] + \log|\mathbf{Z}|, \qquad (5)$$

where $\boldsymbol{\theta}$ is a set of spatial dependence parameters (see Table 1), $\mathbf{Z}$ is a $N \times N$ matrix describing spatial interactions, and $\hat{\mathbf{r}}$ is a $N \times 1$ vector of residuals. Table 1 summarizes definitions $\boldsymbol{\theta}$, $\mathbf{Z}$, and $\hat{\mathbf{r}}$ for SLM, SEM, SAC, and SDM. For all models, the determinant $|\mathbf{Z}|$ and the inverse $\mathbf{Z}^{-1}$ must be evaluated, and its computational complexities are $O(N^3)$, which is intractable for large samples. An objective of this study is to establish how to reduce computational costs here. For existing computationally efficient likelihood approximations, see Smirnov and Anselin (2001), Griffith (2004; 2015), and LeSage and Pace (2009).

**[Table 1 around here]**

## 2.3. Direct effects and indirect effects

The spatial econometric literature reveals that in the SLM, the SDM, or the SAC model, parameter $\boldsymbol{\beta}$ itself is not easy to interpret because it is confounded by the term $\rho \mathbf{W}\mathbf{y}$. Hence the direct effects (DE), $\partial y_i / \partial x_{i,k}$, and indirect effects (IE), $\partial y_j / \partial x_{i,k}$, where $x_{i,k}$ is the *i*-th element of the *k*-th covariate, and $y_i$ is the *i*-th element of **y**, are used to understand marginal effects. DE quantifies the impact of a unit change of $x_{i,k}$ on the *i*-th sample, while IE quantifies the impact of the change on neighboring samples; i.e., IE quantities spatial spill-over effects. DE and IE for SLM/SDM/SAC have $N$ and $N(N - 1)$ realizations, respectively. As a summary measure, their averaged values are usually used. The average DE is defined by $DE_k = \frac{1}{N} \sum_{i=1}^{N} \frac{\partial y_i}{\partial x_{i,k}}$, whereas the average IE is defined by $IE_k = \frac{1}{N} \sum_j \sum_{i \neq j} \frac{\partial y_j}{\partial x_{i,k}}$ (Table 2). For further details about DE and IE, see LeSage and Pace (2009).

[Table 2 around here]

## 3. Low rank spatial econometric models

In this section, we propose low rank spatial econometric models. The first, sub-section 3.1 formulates low rank spatial econometric models. Sub-section 3.2 introduces a Type II restricted likelihood (empirical Bayes) approach to estimate the parameters of these models in a computationally fast manner. Finally, the merit of our proposed model compared to a standard specification is discussed in sub-section 3.6.

### 3.1. Low rank approximations

Let us define the $\mathbf{W}$ matrix as $\frac{1}{\lambda_{max}}\mathbf{W}^{(0)}$, where $\mathbf{W}^{(0)}$ is a symmetric spatial proximity matrix with zero diagonal entries, and $\lambda_{max}$ is the maximum eigenvalue. Then, the $(\mathbf{I} - \varphi\mathbf{W})^{-1}$ matrix may be rewritten as follows:

$$(\mathbf{I} - \varphi\mathbf{W})^{-1} = (\mathbf{E}_*\mathbf{E}_*' - \varphi\mathbf{E}_*\mathbf{\Lambda}_*\mathbf{E}_*')^{-1} = [\mathbf{E}_*(\mathbf{I}_L - \varphi\mathbf{\Lambda}_*)\mathbf{E}_*']^{-1}$$
$$\approx \mathbf{E}_*(\mathbf{I}_L - \varphi\mathbf{\Lambda}_*)^{-1}\mathbf{E}_*',$$
(6)

where $\mathbf{I}_L$ is an $L \times L$ identity matrix, $\mathbf{E}_*$ is a matrix containing all of the $N$ eigenvectors, and $\mathbf{\Lambda}_*$ is a diagonal matrix whose entries are the $N$ eigenvalues, with the eigenvalue and eigenvector entries in the same order. Eq. (6) suggests that $(\mathbf{I} - \varphi\mathbf{W})^{-1} = \mathbf{E}_*(\mathbf{I}_L - \varphi\mathbf{\Lambda}_*)^{-1}\mathbf{E}_*'$ can be approximated by $\mathbf{E}(\mathbf{I}_L - \varphi\mathbf{\Lambda})^{-1}\mathbf{E}'$, where $\mathbf{E}$ is a $N \times L$ matrix of $L$ ($\leq N$) eigenvectors, and $\mathbf{\Lambda}$ is an $L \times L$ diagonal matrix of the corresponding $L$ eigenvalues. The eigenpairs in $\mathbf{E}$ and $\mathbf{\Lambda}$ are assumed to be arranged in descending order. The $l$-th eigenvalue is denoted by $\lambda_l$, where $l \in \{1, \cdots L\}$. The eigenvalues of $\mathbf{W} = \frac{1}{\lambda_{max}} \mathbf{W}^{(0)}$ take a value between 1 ($= \frac{\lambda_{max}}{\lambda_{max}}$) and $\frac{\lambda_{min}}{\lambda_{max}}$, where $\lambda_{min}(= \lambda_L)$ is the minimum eigenvalue of matrix $\mathbf{W}_0$. Accordingly, $\varphi$ is restricted to take a value between 1 and $\frac{\lambda_{min}}{\lambda_{max}}$, just like for the conventional spatial econometric models.

Using this preceding expression, the SEM, Eq. (1) is readily approximated as follows:

$$\mathbf{y} = \mathbf{X}\boldsymbol{\beta} + (\mathbf{I} - \varphi\mathbf{W})^{-1}\boldsymbol{\varepsilon} \approx \mathbf{X}\boldsymbol{\beta} + \mathbf{E}(\mathbf{I}_L - \varphi\boldsymbol{\Lambda})^{-1}\mathbf{E}'\boldsymbol{\varepsilon},$$

$$= \mathbf{X}\boldsymbol{\beta} + \mathbf{E}\boldsymbol{\gamma} \quad \boldsymbol{\gamma} \sim N(\mathbf{0}, \sigma^2(\mathbf{I}_L - \varphi\boldsymbol{\Lambda})^{-2}),$$

(7)

where the first term represents the trend, and the second term represents spatially dependent errors. An interesting finding is that Eq. (7) does not have an independent error process, which is useful to capture data noise. Given that the existence of an independent error process is a premise in geostatistical modeling, this specification weakness might be a crucial flaw in spatial econometric modeling.

The SLM, Eq. (2), can be approximated by replacing $(\mathbf{I} - \rho\mathbf{W})^{-1}$ with $\mathbf{E}(\mathbf{I}_L - \lambda\boldsymbol{\Lambda})^{-1}\mathbf{E}'$ as follows:

$$\mathbf{y} = (\mathbf{I} - \rho\mathbf{W})^{-1}\mathbf{X}\boldsymbol{\beta} + (\mathbf{I} - \rho\mathbf{W})^{-1}\boldsymbol{\varepsilon}$$

$$\approx \mathbf{E}(\mathbf{I}_L - \rho\boldsymbol{\Lambda})^{-1}\mathbf{E}'\mathbf{X}\boldsymbol{\beta} + \mathbf{E}(\mathbf{I}_L - \rho\boldsymbol{\Lambda})^{-1}\mathbf{E}'\boldsymbol{\varepsilon}.$$

(8)

Unfortunately, this specification has a severe disadvantage in that the basic linear regression model (LM) is no longer a special case of Eq. (8). Specifically, when $\rho = 0$, $\mathbf{E}(\mathbf{I}_L - \rho\boldsymbol{\Lambda})^{-1}\mathbf{E}'\mathbf{X}\boldsymbol{\beta} = \mathbf{E}\mathbf{E}'\mathbf{X}\boldsymbol{\beta} \neq \mathbf{X}\boldsymbol{\beta}$ because $\mathbf{E}\mathbf{E}' \neq \mathbf{I}$ after the rank reduction while $\mathbf{E}'\mathbf{E} = \mathbf{I}_L$. Our preliminary analysis demonstrates that Eq. (8) tends to have lower model accuracy than the basic linear model due to this property.

To overcome this difficulty, we expand the SLM following Tiefelsdorf and Griffith (2007) as follows:

$$\begin{aligned}\mathbf{y} &= (\mathbf{I} - \rho\mathbf{W})^{-1}\mathbf{X}\boldsymbol{\beta} + (\mathbf{I} - \rho\mathbf{W})^{-1}\boldsymbol{\varepsilon} \\ &= (\mathbf{I} + \rho\mathbf{W} + \rho^2\mathbf{W}^2 + \cdots)\mathbf{X}\boldsymbol{\beta} + (\mathbf{I} - \rho\mathbf{W})^{-1}\boldsymbol{\varepsilon} \\ &= \mathbf{X}\boldsymbol{\beta} + \rho\mathbf{W}(\mathbf{I} + \rho\mathbf{W} + \rho^2\mathbf{W}^2 + \cdots)\mathbf{X}\boldsymbol{\beta} + (\mathbf{I} - \rho\mathbf{W})^{-1}\boldsymbol{\varepsilon} \\ &= \mathbf{X}\boldsymbol{\beta} + \rho\mathbf{W}(\mathbf{I} - \rho\mathbf{W})^{-1}\mathbf{X}\boldsymbol{\beta} + (\mathbf{I} - \rho\mathbf{W})^{-1}\boldsymbol{\varepsilon}.\end{aligned}$$

(9)

After this expansion, the SLM has the following low rank representation:

$$\begin{aligned}\mathbf{y} &= \mathbf{X}\boldsymbol{\beta} + \rho\mathbf{E}\boldsymbol{\Lambda}\mathbf{E}'\mathbf{E}(\mathbf{I}_L - \rho\boldsymbol{\Lambda})^{-1}\mathbf{E}'\mathbf{X}\boldsymbol{\beta} + \mathbf{E}(\mathbf{I}_L - \rho\boldsymbol{\Lambda})^{-1}\mathbf{E}'\boldsymbol{\varepsilon}, \\ &= \mathbf{X}\boldsymbol{\beta} + \rho\mathbf{E}\boldsymbol{\Lambda}(\mathbf{I}_L - \rho\boldsymbol{\Lambda})^{-1}\mathbf{E}'\mathbf{X}\boldsymbol{\beta} + \mathbf{E}\boldsymbol{\gamma}, \quad \boldsymbol{\gamma} \sim N(\mathbf{0}, \sigma^2(\mathbf{I}_L - \rho\boldsymbol{\Lambda})^{-2}),\end{aligned}$$

(10)

where the first term, $\mathbf{X\beta}$, measures the direct impact, the second term, $\rho\mathbf{E\Lambda}(\mathbf{I}_L - \rho\mathbf{\Lambda})^{-1}\mathbf{E}'\mathbf{X\beta} = \rho\mathbf{E\Lambda E}'\mathbf{E}(\mathbf{I}_L - \rho\mathbf{\Lambda})^{-1}\mathbf{E}'\mathbf{X\beta} \approx \rho\mathbf{W}(\mathbf{I} - \rho\mathbf{W})^{-1}\mathbf{X\beta}$, measures the indirect spillover effects, and the third term, $\mathbf{E\gamma}$, represents spatially autocorrelated errors. Unlike Eq. (8), Eq. (10) becomes the basic LM if $\rho = 0$. Yet, an independent noise process is absent from Eq. (8). Eq. (10) shows that, as suggested by Arbia (2006), the SLM does not have a representation as a pure spatial process, but has a representation as the sum of a spatial process term, $\mathbf{E\gamma}$, where $\mathbf{\gamma}$ is a vector of random coefficients appearing in Eq.(10), and a deterministic term, $\rho\mathbf{E\Lambda}(\mathbf{I}_L - \rho\mathbf{\Lambda})^{-1}\mathbf{E}'\mathbf{X\beta}$, accounting for the spill-overs from $\mathbf{X}$.

The SDM can be approximated in the same way by replacing $\mathbf{X\beta}$ in Eq. (9) with $\mathbf{X\beta} + \mathbf{WX\gamma}$. The approximate SDM is obtained as follows:

$$\mathbf{y} = \mathbf{X\beta} + \mathbf{WXq} + \rho\mathbf{E\Lambda}(\mathbf{I}_L - \rho\mathbf{\Lambda})^{-1}\mathbf{E}'(\mathbf{X\beta} + \mathbf{WXq}) + \mathbf{E\gamma},$$
$$\mathbf{\gamma} \sim N(\mathbf{0}, \sigma^2(\mathbf{I}_L - \rho\mathbf{\Lambda})^{-2}).$$
(11)

We do not assume any rank reduction for the $\mathbf{W}$ matrix in $\mathbf{WX}$. This is because the computational burden for $\mathbf{WX}$ is very small, even without rank reduction.

The SAC is approximated in the similar way to that for the SLM. Expanding SAC as

$$\mathbf{y} = (\mathbf{I} - \rho\mathbf{W})^{-1}\mathbf{X}\boldsymbol{\beta} + (\mathbf{I} - \rho\mathbf{W})^{-1}(\mathbf{I} - \lambda\mathbf{W})^{-1}\boldsymbol{\varepsilon}$$

$$= (\mathbf{I} + \rho\mathbf{W} + \rho^2\mathbf{W}^2 + \cdots)\mathbf{X}\boldsymbol{\beta} + (\mathbf{I} - \rho\mathbf{W})^{-1}(\mathbf{I} - \lambda\mathbf{W})^{-1}\boldsymbol{\varepsilon}$$

$$= \mathbf{X}\boldsymbol{\beta} + \rho\mathbf{W}(\mathbf{I} + \rho\mathbf{W} + \rho^2\mathbf{W}^2 + \cdots)\mathbf{X}\boldsymbol{\beta} \qquad (12)$$

$$+ (\mathbf{I} - \rho\mathbf{W})^{-1}(\mathbf{I} - \lambda\mathbf{W})^{-1}\boldsymbol{\varepsilon}$$

$$= \mathbf{X}\boldsymbol{\beta} + \rho\mathbf{W}(\mathbf{I} - \rho\mathbf{W})^{-1}\mathbf{X}\boldsymbol{\beta} + (\mathbf{I} - \rho\mathbf{W})^{-1}(\mathbf{I} - \lambda\mathbf{W})^{-1}\boldsymbol{\varepsilon}$$

allows Eq. (12) to be approximated just like the SLM:

$$\mathbf{y} = \mathbf{X}\boldsymbol{\beta} + \rho\mathbf{E}\boldsymbol{\Lambda}\mathbf{E}'\mathbf{E}(\mathbf{I}_L - \rho\boldsymbol{\Lambda})^{-1}\mathbf{E}'\mathbf{X}\boldsymbol{\beta} + \mathbf{E}(\mathbf{I}_L - \rho\boldsymbol{\Lambda})^{-1}\mathbf{E}'\mathbf{E}(\mathbf{I}_L - \lambda\boldsymbol{\Lambda})^{-1}\mathbf{E}'\boldsymbol{\varepsilon},$$
$$(13)$$
$$= \mathbf{X}\boldsymbol{\beta} + \rho\mathbf{E}\boldsymbol{\Lambda}(\mathbf{I}_L - \rho\boldsymbol{\Lambda})^{-1}\mathbf{E}'\mathbf{X}\boldsymbol{\beta} + \mathbf{E}\boldsymbol{\gamma}, \quad \boldsymbol{\gamma} \sim N(\mathbf{0}, \sigma^2(\mathbf{I}_L - \rho\boldsymbol{\Lambda})^{-2}(\mathbf{I}_L - \lambda\boldsymbol{\Lambda})^{-2}).$$

This section shows that the SEM, SLM, SDM, and SAC have low rank representations. At the same time, these results highlight that these models do not assume independent noise process in **y**. Unfortunately, this assumption is unsuitable for noisy data, such as remotely sensed images, small area data (e.g., district-level socio-economic/demographic data), and smart sensor data (e.g., electricity usage) that are now becoming available (see Arbia et al., 2016).

### 3.2. Low rank spatial econometric models

To include a white noise process, we introduce a data model, which is usually assumed in geostatistics (see Section 1). Our data model is specified as

$$\mathbf{y} = \beta_1 \mathbf{1} + \mathbf{z} + \mathbf{u}, \qquad \mathbf{u} \sim N(\mathbf{0}, \tau^2 \mathbf{I}), \tag{14}$$

where $\mathbf{u}$ is a white noise process with variance $\tau^2$, which often is called a nugget parameter in geostatistics. Note that Bivand et al. (2015) also introduce a white noise process into (full rank) spatial econometric models to estimate the model in an integrated nested Laplace approximation framework. $\beta_1$ represents an intercept, and $\mathbf{1}$ is a N-by-1 vector of ones. An intercept is included in the data model because it increases identifiability (see the subsequent discussion).

This study specifies the process $\mathbf{z}$, which can be considered as a vector of de-noised observations, by a spatial econometric model without an intercept term. The low rank SEM (LSEM) is defined by substituting Eq. (7), which approximates the SEM, into $\mathbf{z}$ as follows:

$$\mathbf{y} = \mathbf{X}\boldsymbol{\beta} + \mathbf{E}\boldsymbol{\gamma} + \mathbf{u}, \qquad \boldsymbol{\gamma} \sim N(\mathbf{0}, \sigma^2(\mathbf{I}_L - \varphi\boldsymbol{\Lambda})^{-2}), \qquad \mathbf{u} \sim N(\mathbf{0}, \tau^2 \mathbf{I}). \tag{15}$$

where $(\mathbf{I}_L - \varphi\boldsymbol{\Lambda})^{-2}$ is a diagonal matrix whose *l*-th entry equals $1/(1 - \varphi\lambda_l)^2$. If $\varphi$ is large (i.e., near 1), the diagonal elements corresponding to large eigenvalues are considerably greater than the diagonal elements corresponding to small eigenvalues. As a result, the coefficients corresponding to principal eigenvectors, which explain large-scale map patterns, become large. As a result, the error process has a large-scale map pattern. By contrast, as $\varphi$ approaches zero, the decay becomes slow, and the resulting process has small-scale pattern. A notable difference with the original SEM is that the error process disappears not when $\varphi = 0$, but when $\sigma^2 = 0$. In other words, $\varphi$ in the LSEM is purely a scale parameter, and $\sigma^2$ is an intensity parameter. The parameters $\{\varphi, \sigma^2, \tau^2\}$ correspond to the range, partial-sill, and nugget parameters in geostatistics.

The low rank SLM (LSLM) is specified by substituting Eq. (10), with an intercept of zero, into $\mathbf{z}$, as

$$\mathbf{y} = \beta_1 \mathbf{1} + [\mathbf{I} + \rho \mathbf{E} \Lambda (\mathbf{I}_L - \rho \Lambda)^{-1} \mathbf{E}'] \mathbf{X}_{-1} \boldsymbol{\beta}_{-1} + \mathbf{E} \boldsymbol{\gamma} + \mathbf{u}$$

(16)

$$\boldsymbol{\gamma} \sim N(\mathbf{0}, \sigma^2 (\mathbf{I}_L - \rho \Lambda)^{-2}), \qquad \mathbf{u} \sim N(\mathbf{0}, \tau^2 \mathbf{I}),$$

where $\mathbf{X} = [\mathbf{1}, \mathbf{X}_{-1}]$ and $\boldsymbol{\beta} = [\beta_1, \boldsymbol{\beta}'_{-1}]'$. Note that the model may be defined as $\mathbf{y} = [\mathbf{I} + \rho \mathbf{E} \Lambda (\mathbf{I}_L - \rho \Lambda)^{-1} \mathbf{E}'] \mathbf{X} \boldsymbol{\beta} + \mathbf{E} \boldsymbol{\gamma} + \mathbf{u}$, which is more consistent with the standard SLM. However, our preliminary analysis shows that, in this specification, $\beta_1 \rho \mathbf{E} \Lambda (\mathbf{I}_L - \rho \Lambda)^{-1} \mathbf{E}' \mathbf{1}$ in the first term and $\mathbf{E} \boldsymbol{\gamma}$ tend to be collinear, and the identification of the true intercept value becomes difficult.[2] Therefore, we define the data model as Eq. (14), and the LSLM as Eq. (16). While the standard SLM captures spatial interactions among observations (see Manski, 1993), the LSLM captures the interactions among denoised observations $\mathbf{z}$, using the term $[\mathbf{I} + \rho \mathbf{E}\boldsymbol{\Lambda}(\mathbf{I}_L - \rho\boldsymbol{\Lambda})^{-1}\mathbf{E}']\mathbf{X}_{-1}\boldsymbol{\beta}_{-1} + \mathbf{E}\boldsymbol{\gamma}$ to approximate $(\mathbf{I} - \rho\mathbf{W})^{-1}(\mathbf{X}_{-1}\boldsymbol{\beta}_{-1} + \boldsymbol{\varepsilon}) = (\mathbf{I} - \rho\mathbf{W})^{-1}\mathbf{z}$. Bivand et al. (2015) introduce a similar SLM specification.

The low rank SDM (LSDM) is obtained in the same way as the LSLM:

$$\mathbf{y} = \beta_1 \mathbf{1} + [\mathbf{I} + \rho \mathbf{E}\boldsymbol{\Lambda}(\mathbf{I}_L - \rho\boldsymbol{\Lambda})^{-1}\mathbf{E}'](\mathbf{X}_{-1}\boldsymbol{\beta}_{-1} + \mathbf{W}\mathbf{X}_{-1}\mathbf{q}_{-1}) + \mathbf{E}\boldsymbol{\gamma} + \mathbf{u} \tag{17}$$

$$\boldsymbol{\gamma} \sim N(\mathbf{0}, \sigma^2(\mathbf{I}_L - \rho\boldsymbol{\Lambda})^{-2}), \qquad \mathbf{u} \sim N(\mathbf{0}, \tau^2\mathbf{I}),$$

where $\mathbf{q} = [q_1, \mathbf{q}'_{-1}]'$.

Finally, the low rank SAC (LSAC) is defined as

$$\mathbf{y} = \beta_1 \mathbf{1} + [\mathbf{I} + \rho \mathbf{E}\boldsymbol{\Lambda}(\mathbf{I}_L - \rho\boldsymbol{\Lambda})^{-1}\mathbf{E}']\mathbf{X}_{-1}\boldsymbol{\beta}_{-1} + \mathbf{E}\boldsymbol{\gamma} + \mathbf{u} \tag{18}$$

$$\boldsymbol{\gamma} \sim N(\mathbf{0}, \sigma^2(\mathbf{I}_L - \rho\boldsymbol{\Lambda})^{-2}(\mathbf{I}_L - \varphi\boldsymbol{\Lambda})^{-2}), \qquad \mathbf{u} \sim N(\mathbf{0}, \tau^2\mathbf{I}).$$

Given Eq. (14), the three models include a white noise process, $\mathbf{u}$; they are more suitable for noisy data than are the classical SEM, SLM, SDM, and SAC specifications. Besides, the low rank models have hierarchical representations as summarized in Table 3. In other words, the low rank models are expressed as $p(\mathbf{y}|\mathbf{z})p(\mathbf{z}|\boldsymbol{\gamma})p(\boldsymbol{\gamma}) =$ [data | spatial process][spatial process | parameter][parameter], which typically is assumed in recent geostatistical studies.

---

[2] This might be the reason why the intercept of the classic SLM often takes a singular value.

[Table 3 around here]

One issue concerns how to select the number of eigenvectors, *L*. A large *L* decreases model errors, but increases model complexity that can lead to overfitting. Thus, *L* must be selected to balance the accuracy and complexity of the model. *L* can be given by the number of eigen-pairs satisfying $|\lambda_l/\lambda_{\max}| > t$, where *t* is a threshold value. Following Griffith (2003), this threshold can be 0.25, 0.50, or 0.75. Alternatively, *L* can be determined adaptively, considering the data size and the computing environment. This is because a large *L* increases computational cost. Based on the simulation experiment in Section 4, $L \geq 200$ is desirable.

Because $[\mathbf{I} + \rho\mathbf{E}\boldsymbol{\Lambda}(\mathbf{I}_L - \rho\boldsymbol{\Lambda})^{-1}\mathbf{E}']$ approximates $(\mathbf{I} - \rho\mathbf{W})^{-1}$, the direct and indirect effects (see Table 2) for the LSLM and LSAC can be approximated by the diagonal and off-diagonal elements of $[\mathbf{I} + \rho\mathbf{E}\boldsymbol{\Lambda}(\mathbf{I}_L - \rho\boldsymbol{\Lambda})^{-1}\mathbf{E}']\widehat{\boldsymbol{\beta}}_{-1}$, respectively. $DE_k$ may be

defined as follows:

$$DE_k = \frac{1}{N} Tr[\mathbf{I} + \rho \mathbf{E} \mathbf{\Lambda} (\mathbf{I}_L - \rho \mathbf{\Lambda})^{-1} \mathbf{E}'] \hat{\beta}_k$$

$$= \frac{1}{N} \sum_{i=1}^{N} [1 + \rho \tilde{\mathbf{e}}'_i \mathbf{\Lambda} (\mathbf{I}_L - \rho \mathbf{\Lambda})^{-1} \tilde{\mathbf{e}}_i] \hat{\beta}_k, \quad (19)$$

where $\tilde{\mathbf{e}}_i$ is the *i*-th row of $\mathbf{E}$. $Tr[\mathbf{B}]$ returns the trace of the matrix $\mathbf{B}$. In contrast, $IE_k$ may be defined as follows:

$$IE_k = \frac{1}{N} \mathbf{1}'[\mathbf{I} + \rho \mathbf{E} \mathbf{\Lambda} (\mathbf{I}_L - \rho \mathbf{\Lambda})^{-1} \mathbf{E}'] \mathbf{1} \hat{\beta}_k - DE_k$$

$$= \hat{\beta}_k + \hat{\beta}_k \frac{\rho}{N} \mathbf{1}' \mathbf{E} \mathbf{\Lambda} (\mathbf{I}_L - \rho \mathbf{\Lambda})^{-1} \mathbf{E}' \mathbf{1} - DE_k \quad (20)$$

A notable feature of this specification is that it does not need to store a $N \times N$ matrix.

In the case of the LSDM, $DE_k$ and $IE_k$ may be defined as

$$DE_k = \frac{1}{N} Tr[(\mathbf{I} + \rho \mathbf{E} \mathbf{\Lambda} (\mathbf{I}_L - \rho \mathbf{\Lambda})^{-1} \mathbf{E}')(\hat{\beta}_k \mathbf{I} + \hat{q}_k \mathbf{W})],$$

$$= \frac{1}{N} \sum_{i=1}^{N} [1 + \rho \tilde{\mathbf{e}}'_i \mathbf{\Lambda} (\mathbf{I}_L - \rho \mathbf{\Lambda})^{-1} \tilde{\mathbf{e}}_i] \hat{\beta}_k + \frac{1}{N} \sum_{i=1}^{N} [\rho \tilde{\mathbf{e}}'_i \mathbf{\Lambda} (\mathbf{I}_L - \rho \mathbf{\Lambda})^{-1} \tilde{\mathbf{e}}_i^{(\mathbf{W})}] \hat{q}_k, \quad (21)$$

$$IE_k = \frac{1}{N} \mathbf{1}'[\mathbf{I} + \rho \mathbf{E} \mathbf{\Lambda} (\mathbf{I}_L - \rho \mathbf{\Lambda})^{-1} \mathbf{E}'] (\mathbf{1} \hat{\beta}_k + \mathbf{W} \mathbf{1} \hat{q}_k) - DE_k,$$

$$= \hat{\beta}_k + \frac{\rho}{N} \mathbf{1}' \mathbf{E} \mathbf{\Lambda} (\mathbf{I}_L - \rho \mathbf{\Lambda})^{-1} \mathbf{E}' (\mathbf{1} \hat{\beta}_k + \mathbf{W} \mathbf{1} \hat{q}_k) + \frac{\mathbf{1}' \mathbf{W} \mathbf{1}}{N} \hat{q}_k - DE_k, \quad (22)$$

where $\tilde{\mathbf{e}}_i^{(\mathbf{W})}$ is the *i*-th row of the $\mathbf{WE}$ matrix.

### 3.3. Parameter estimation

Spatial econometric models typically are estimated by maximizing the likelihood $p(\mathbf{y}|\boldsymbol{\theta})$, which involves a determinant evaluation and an inversion of $\mathbf{Z}$, both of which have complexy of $O(N^3)$, where $\boldsymbol{\theta}$ represents spatial dependence parameters, and $\mathbf{Z}$ equals $(\mathbf{I}_L - \rho \boldsymbol{\Lambda})^{-2}$ for the SLM (see Table 1). However, here we maximize the Type II (empirical Bayes/$h$-) likelihood, which in our case is formulated as $p(\mathbf{y}|\boldsymbol{\gamma}, \boldsymbol{\theta})p(\boldsymbol{\gamma}|\boldsymbol{\theta})$ because it is more consistent with our hierarchical representation, and, more importantly, it allows us to estimate parameters in a more computationally efficient manner.

Before this estimation, we rewrite the low rank models as follows:

$$\mathbf{y} = \mathbf{X}_{\boldsymbol{\theta}}\boldsymbol{\beta} + \mathbf{E}\boldsymbol{\Sigma}_{\boldsymbol{\theta}}\mathbf{v} + \mathbf{u} \qquad \mathbf{v} \sim N(\mathbf{0}, \tau^2 \mathbf{I}_L) \qquad \mathbf{u} \sim N(\mathbf{0}, \tau^2 \mathbf{I}), \qquad (23)$$

where $\boldsymbol{\Sigma}_{\boldsymbol{\theta}} \mathbf{v} = \boldsymbol{\gamma}$; $\mathbf{X}_{\boldsymbol{\theta}}$ and $\boldsymbol{\Sigma}_{\boldsymbol{\theta}}$ are summarized in Table 4.

**[Table 4 around here]**

Eq. (23) is identical to the linear mixed effects (LME) model (e.g., Bates, 2010) Two types of likelihood are available. The Type I likelihood, $l_1(\mathbf{y})$, is evaluated using the probability density function (PDF) of $\mathbf{y} \sim N(\mathbf{X_\theta \beta}, \tau^2 \mathbf{E \Sigma_\theta^2 E'} + \tau^2 \mathbf{I})$, which is identical to Eq. (23). The Type II likelihood is defined by $l_2(\mathbf{y}) = \int p(\mathbf{y}|\mathbf{v})p(\mathbf{v}) \, d\mathbf{v}$, where $p(\mathbf{y}|\mathbf{v})$ and $p(\mathbf{v})$ are PDFs for $\mathbf{y} \sim N(\mathbf{X_\theta \beta} + \mathbf{E\Sigma_\theta v}, \tau^2 \mathbf{I})$ and $\mathbf{v} \sim N(\mathbf{0}, \tau^2 \mathbf{I}_L)$, respectively (e.g., Bates, 2010). Mainly for small area estimation, Salvati (2004), Pratesi and Salvati (2009), among others, have estimated LME models with (full rank) spatially dependent errors using Type I likelihood maximization. To the best of the authors' knowledge, Type II likelihood maximization has never been used in spatial econometric modeling. However, based on Pinheiro and Bates (2000), the second specification is computationally more efficient. Besides, because Type II likelihood maximization is identical to empirical Bayesian estimation, it is readily extended to a wider class of Bayesian spatial models. We prefer the Type II approach.

We maximize the Type II restricted log-likelihood, $loglik_R(\mathbf{\theta})$, following Murakami and Griffith (2015; 2018a), who estimate a random effects ESF model. The estimation procedure is as follows:

(i)     $\mathbf{\theta}$ is estimated by numerically maximizing Eq.(24) with plugins of Eqs. (25) and (26),

(ii)    $\{\widehat{\boldsymbol{\beta}}, \widehat{\mathbf{v}}\}$ are estimated by Eq. (26),

(iii)   $\hat{\tau}^2$ is estimated using Eq. (27).

$$loglik_R(\boldsymbol{\theta}) = -\frac{1}{2}ln\left|\begin{bmatrix}\mathbf{X'_\theta X_\theta} & \mathbf{X'_\theta E\Sigma_\theta} \\ \mathbf{\Sigma_\theta E'X_\theta} & \mathbf{\Sigma_\theta^{-2} + I_L}\end{bmatrix}\right| - \frac{N-K}{2}\left(1 + ln\left(\frac{2\pi d(\boldsymbol{\theta})}{N-K}\right)\right), \quad (24)$$

$$d(\boldsymbol{\theta}) = (\mathbf{y} - \mathbf{X_\theta}\widehat{\boldsymbol{\beta}} - \mathbf{E\Sigma_\theta}\widehat{\mathbf{v}})'(\mathbf{y} - \mathbf{X_\theta}\widehat{\boldsymbol{\beta}} - \mathbf{E\Sigma_\theta}\widehat{\mathbf{v}}) + \widehat{\mathbf{v}}'\widehat{\mathbf{v}} \quad (25)$$

$$\begin{bmatrix}\widehat{\boldsymbol{\beta}} \\ \widehat{\mathbf{v}}\end{bmatrix} = \begin{bmatrix}\mathbf{X'_\theta X_\theta} & \mathbf{X'_\theta E\Sigma_\theta} \\ \mathbf{\Sigma_\theta E'X_\theta} & \mathbf{\Sigma_\theta^2 + I_L}\end{bmatrix}^{-1}\begin{bmatrix}\mathbf{X'_\theta y} \\ \mathbf{\Sigma_\theta E'y}\end{bmatrix} \quad (26)$$

$$\hat{\tau}^2 = \frac{1}{N-K}(\mathbf{y} - \mathbf{X_\theta}\boldsymbol{\beta} - \mathbf{E\Sigma_\theta v})'(\mathbf{y} - \mathbf{X_\theta}\boldsymbol{\beta} - \mathbf{E\Sigma_\theta v}) \quad (27)$$

Eq. (25) balances accuracy and complexity of a model. For further detail about this estimation approach, see Bates (2010) and Murakami and Griffith (2015).

The variance-covariance matrix for the estimated coefficients is given by

$$Var\begin{bmatrix}\widehat{\boldsymbol{\beta}} \\ \widehat{\mathbf{v}}\end{bmatrix} = \hat{\tau}^2\begin{bmatrix}\mathbf{X'_\theta X_\theta} & \mathbf{X'_\theta E\Sigma_\theta} \\ \mathbf{\Sigma_\theta E'X_\theta} & \mathbf{\Sigma_\theta^2 + I_L}\end{bmatrix}^{-1}. \quad (28)$$

The diagonal entries are useful to evaluate the statistical significance of the estimated coefficients.

### 3.4. Bootstrapping

This section introduces a parametric bootstrap that estimates confidence intervals for the spatial dependence parameters (see Table 1), $DE_k$ and $IE_k$. Because Eq. (23) is identical to the usual LME model, bootstrapping for the standard LME model already is available. The procedure may be summarized as follows:

(i)     $\mathbf{v} \sim N(\mathbf{0}, \hat{\tau}^2 \mathbf{I}_L)$ and $\mathbf{u} \sim N(\mathbf{0}, \hat{\tau}^2 \mathbf{I})$ are randomly sampled;

(ii)    $\mathbf{y}^* = \mathbf{X}_{\hat{\boldsymbol{\theta}}} \widehat{\boldsymbol{\beta}} + \mathbf{E} \boldsymbol{\Sigma}_{\hat{\boldsymbol{\theta}}} \mathbf{v} + \mathbf{u}$ is calculated;

(iii)   Parameters and direct/indirect effects are estimated by applying the restricted log-likelihood maximization (REML), which we introduced in Section 3.3, to a low rank spatial econometric model whose $\mathbf{y}$ is replaced with $\mathbf{y}^*$;

(iv)    Steps (i) to (iv) are iterated *m* times.

**3.5. Fast computations**

Our model estimation involves an eigen-decomposition, parameter estimation, and bootstrapping. Unfortunately, all of these tasks can be time consuming for large samples. This section explains, in the subsequent three subsections, strategies for reducing the computational costs for these three tasks.

3.5.1. Fast eigen-decomposition

The eigen-decomposition, whose computational complexity equals $O(N^3)$, is prohibitive for large samples. For a sparse matrix $\mathbf{W}$, the ARPACK (https://www.caam.rice.edu/software/ARPACK/) or other fast eigen-decomposition routines are useful to reduce computational burden (see Griffith, 2000). For a dense $\mathbf{W}$, the Nyström approximation (Drineas and Mahoney, 2005) or other eigen-approximations, which appear in the machine learning literature for eigen-decomposition of kernel matrices, are useful.

3.5.2. Fast parameter estimation

A key feature of our specification is that matrices and vectors whose sizes depend on $N$ can be eliminated before executing estimation and bootstrapping. Although we consider the LSEM as an example, similar approaches are available for the LSLM, LSDM, and LSAC (see Appendix.1). The restricted log-likelihood (Eqs. (24)-(26)) for the LSEM can be rewritten as follows:

$$loglik_R(\boldsymbol{\theta}) = -\frac{1}{2} ln \left| \begin{bmatrix} \mathbf{M}_{XX} & \mathbf{M}_{EX}\boldsymbol{\Sigma}_{\boldsymbol{\theta}} \\ \boldsymbol{\Sigma}_{\boldsymbol{\theta}}\mathbf{M}'_{EX} & \boldsymbol{\Sigma}_{\boldsymbol{\theta}}^2 + \mathbf{I}_L \end{bmatrix} \right| - \frac{N-K}{2}\left(1 + log\left(\frac{2\pi d(\boldsymbol{\theta})}{N-K}\right)\right), \quad (29)$$

$$d(\boldsymbol{\theta}) = m_{yy} - 2[\widehat{\boldsymbol{\beta}}', \widehat{\mathbf{v}}'] \begin{bmatrix} \mathbf{m}_{Xy} \\ \boldsymbol{\Sigma}_{\boldsymbol{\theta}} \mathbf{m}_{Ey} \end{bmatrix} + [\widehat{\boldsymbol{\beta}}', \widehat{\mathbf{v}}'] \begin{bmatrix} \mathbf{M}_{XX} & \mathbf{M}_{EX} \boldsymbol{\Sigma}_{\boldsymbol{\theta}} \\ \boldsymbol{\Sigma}_{\boldsymbol{\theta}} \mathbf{M}'_{EX} & \boldsymbol{\Sigma}_{\boldsymbol{\theta}}^2 \end{bmatrix} \begin{bmatrix} \widehat{\boldsymbol{\beta}} \\ \widehat{\mathbf{v}} \end{bmatrix} + \widehat{\mathbf{v}}'\widehat{\mathbf{v}}, \quad (30)$$

$$\begin{bmatrix} \widehat{\boldsymbol{\beta}} \\ \widehat{\mathbf{v}} \end{bmatrix} = \begin{bmatrix} \mathbf{M}_{XX} & \mathbf{M}_{EX} \boldsymbol{\Sigma}_{\boldsymbol{\theta}} \\ \boldsymbol{\Sigma}_{\boldsymbol{\theta}} \mathbf{M}'_{EX} & \boldsymbol{\Sigma}_{\boldsymbol{\theta}}^2 + \mathbf{I}_L \end{bmatrix}^{-1} \begin{bmatrix} \mathbf{m}_{Xy} \\ \boldsymbol{\Sigma}_{\boldsymbol{\theta}} \mathbf{m}_{Ey} \end{bmatrix}, \quad (31)$$

where $\mathbf{M}_{XX} = \mathbf{X}'\mathbf{X}$, $\mathbf{M}_{EX} = \mathbf{E}'\mathbf{X}$, $\mathbf{m}_{Xy} = \mathbf{X}'\mathbf{y}$, $\mathbf{m}_{Ey} = \mathbf{E}'\mathbf{y}$, and $m_{yy} = \mathbf{y}'\mathbf{y}$. Interestingly, Eqs. (29) – (31) do not include any matrix or vector with a dimension depending on $N$. The computationally demanding parts in the $loglik_R(\boldsymbol{\theta})$ expression are the inverse and the determinant evaluation of the $\begin{bmatrix} \mathbf{M}_{XX} & \mathbf{M}_{EX} \boldsymbol{\Sigma}_{\boldsymbol{\theta}} \\ \boldsymbol{\Sigma}_{\boldsymbol{\theta}} \mathbf{M}'_{EX} & \boldsymbol{\Sigma}_{\boldsymbol{\theta}}^2 + \mathbf{I}_L \end{bmatrix}$ matrix. Both of these operations have complexities of $O((K+L)^3)$. Thus, if only the five inner products are evaluated a priori, the computational cost for the numerical maximization of the likelihood function is quite small, and independent of the sample size $N$. Murakami and Seya (2017) apply the same idea to a spatial unconditional quantile regression, whereas Murakami and Griffith (2018b) apply it to a spatially varying coefficients model for large samples.

### 3.5.3. Fast bootstrapping

The bootstrap method also can be accelerated by employing a similar idea. In the case of the LSEM, $\mathbf{X}$ and $\mathbf{E}$ are unchanged across iterations. In other words, the only elements we need to calculate in each iteration are $\mathbf{m}_{Xy} = \mathbf{X}'\mathbf{y}$, $\mathbf{m}_{Ey} = \mathbf{E}'\mathbf{y}$, and $m_{yy} =$

$\mathbf{y}'\mathbf{y}$. Fortunately, their total number of operations is only $(K + L + 1)N$. The computational cost is quite small even for large samples. The same approach is available for the LSLM, LSDM, and LSAC (see Appendix.1).

Finally, to estimate the standard errors for $DE_k$ and $IE_k$, which are required to evaluate their statistical significances, we need to evaluate these two quantities in each iteration. With regard to $DE_k$, Eq. (19) must be calculated repeatedly for the LSLM/LSAC ratio, while Eq. (21) must be calculated repeatedly for the LSDM. During these calculations, $\tilde{\mathbf{e}}'_i \Lambda (\mathbf{I}_L - \rho \Lambda)^{-1} \tilde{\mathbf{e}}_i$ and $\tilde{\mathbf{e}}'_i \Lambda (\mathbf{I}_L - \rho \Lambda)^{-1} \tilde{\mathbf{e}}_i^{(\mathbf{W})}$ (only for LSDM) must be evaluated repeatedly. Fortunately, because $\Lambda (\mathbf{I}_L - \rho \Lambda)^{-1}$ is a diagonal matrix, and $\tilde{\mathbf{e}}_i$ and $\tilde{\mathbf{e}}_i^{(\mathbf{W})}$ are short vectors ($L \times 1$), the computational costs of these computations are trivial.

For the LSLM and LSAC, the cost for $IE_k$ is also trivial after $\mathbf{m}_{1E} = \mathbf{1}'\mathbf{E}$ is evaluated. Given $\mathbf{m}_{1E}$, $IE_k$ yields

$$IE_k = \hat{\beta}_k + \hat{\beta}_k \frac{\rho}{N} \mathbf{m}_{1E} \Lambda (\mathbf{I}_L - \rho \Lambda)^{-1} \mathbf{m}'_{1E} - DE_k, \qquad (32)$$

whose computational complexity is independent of sample size. In the case of the LSDM, we additionally need to evaluate $m_W = \mathbf{1}'\mathbf{W}\mathbf{1}$ and $\mathbf{m}_{EW1} = \mathbf{E}\mathbf{W}\mathbf{1}$. Given these elements, the $IE_k$ quantity for the LSDM can be expressed as

$$IE_k = \hat{\beta}_k + \frac{\rho}{N}\mathbf{m}_{1E}\mathbf{\Lambda}(\mathbf{I}_L - \rho\mathbf{\Lambda})^{-1}(\mathbf{m}'_{1E}\hat{\beta}_k + \mathbf{m}_{EW1}\hat{q}_k) + \frac{m_W}{N}\hat{q}_k - DE_k, \qquad (33)$$

which again does not involve any matrix or vectors whose size depends on *N*. Thus, both *DE*$_k$ and *IE*$_k$ can be evaluated computationally efficiently in bootstrapping.

### 3.6. Usefulness of hierarchical low rank specifications

Our specification has at least three specific advantages over the standard spatial econometric model specifications. The first advantage is computational efficiency. Unlike the standard ML estimation for spatial econometric models, our likelihood maximization does not involve inverse and determinant evaluation of an $N \times N$ matrix (the **Z** matrix in Table 1) owing to rank reduction.

The second advantage is expandability. Our models are identical to the LME model (see Bates, 2010) that has been used to estimate multilevel, non-linear, and spatially and non-spatially varying effects, among others. These effects are readily incorporated into our model by adding a term **Gr**, where **G** is a matrix describing these effects (e.g., in the case of multilevel effects, the (*i*, *g*)-th column contains 1 if the *i*-th sample is in the *g*-th group, and 0 otherwise), and **r** is a vector of random coefficients. Besides, our model specifications are akin to the hierarchical geostatistical model specification that has been extended to dynamic spatiotemporal modeling, multivariate spatial modeling, data fusion, and many other contexts (see, Gelfand et al., 2010).

The third advantage of our specification is its explicit consideration of data noise, as we explain in our preceding discussion. The next section shows that consideration of data noise significantly changes analysis results.

## 4. Monte Carlo Simulation Experiment

In this section, we focus on SLM and SEM-type specifications, the two most common specifications employed in spatial econometrics. A simulation experiment focusing on the SAC model or the SDM would be an important future study.
.

### 4.1. Simulation experiment 1

This section compares estimation accuracy of the classical LM, the LSLM, and the SLM while assuming the SLM to be the true data generating process. Section 4.1.1 summarizes our simulation experimental design, and Section 4.1.2 explains the simulation result.

### 4.1.1. The setting

This section compares the LSLM and SLM by fitting them with simulated data generated by Eq. (34):

$$\mathbf{y} = \beta_0 \mathbf{1} + \mathbf{z} + \mathbf{u} \qquad \mathbf{u} \sim N(\mathbf{0}, \tau^2 \mathbf{I}),$$

$$\mathbf{z} = (\mathbf{I} - \rho \mathbf{W})^{-1}(\beta_1 \mathbf{x}_1 + \beta_2 \mathbf{x}_2) + (\mathbf{I} - \rho \mathbf{W})^{-1} \boldsymbol{\varepsilon} \qquad \boldsymbol{\varepsilon} \sim N(\mathbf{0}, \mathbf{I}).$$

(34)

We assume that $\mathbf{x}_1$ is an influential variable, whereas $\mathbf{x}_2$ is a less influential variable. Based on these assumptions, the true coefficient values are as follows: $\beta_0 = 1.0$, $\beta_1 = 2.0$, and $\beta_2 = 0.5$. Following LeSage and Pace (2014), Arbia et al. (2016) among others, covariates $\mathbf{x}_1$ and $\mathbf{x}_2$ are independently sampled from normal distributions, $N(\mathbf{0}, \mathbf{I})$. $\mathbf{W} = \frac{1}{\lambda_{max}} \mathbf{W}_0$, where $\mathbf{W}_0$ is given by a matrix with its $(i,j)$-th element equal to 1 if the samples $i$ and $j$ share more than one border, and 0 if they do not share any border. Spatial coordinates for each sample are generated from two independent standard normal distributions, and the borders are drawn using a Delaunay triangulation. Unlike the conventional SLM, Eq. (34) includes the white noise process $\mathbf{u}$. In other words, the SLM implicitly assumes $\tau^2 = 0$.

In the simulation experiment, synthetic data are replicated 200 times in each case, with $N \in \{500, 2000\}$, $\rho \in \{0.2, 0.4, 0.6, 0.8\}$, and $\tau^2 \in \{0, 2, 4\}$.

The models being compared include the LM, LSLM$_L$, LSLM$_{50}$, LSLM$_{100}$, LSLM$_{200}$, LSLM$_{300}$, LSLM$_{400}$, and SLM, where LSLM$_L$ denotes the LSLM with $L$ eigen-pairs. The eigen-pairs are extracted using rARPACK (https://cran.r-project.org/web/packages/rARPACK/index.html), which is an R package that implements the ARPACK procedure for efficiently calculating eigen-pairs of a sparse matrix.

Estimation accuracies are compared using the root mean squared error (RMSE) and the mean bias. In terms of the $k$-th regression coefficient, $\beta_k$, where $k \in \{1,2,3\}$, these diagnostics are defined as

$$RMSE[\beta_k] = \sqrt{\frac{1}{200} \sum_{iter=1}^{200} (\hat{\beta}_k^{(iter)} - \beta_k)^2} \qquad (35)$$

$$Bias[\beta_k] = \frac{1}{200} \sum_{iter=1}^{200} (\hat{\beta}_k^{(iter)} - \beta_k) \qquad (36)$$

where $\hat{\beta}_k^{(iter)}$ is the estimate in the *iter*-th iteration.

### 4.1.2. Results

Because results for the estimated coefficients $\hat{\beta}_1$ and $\hat{\beta}_2$ are similar, we report results only for $\hat{\beta}_1$. Table 5 summarizes RMSE and bias of the estimates. As expected, the RMSEs and biases of the LM estimates are large when $\rho$ is large. The RMSEs and biases for the SLM and LSLM tend to be smaller than for the LM estimates. If the noise variance $\tau^2$ equals zero (i.e., data is clean), the SLM estimates are especially accurate. However, the RMSEs and biases rapidly increase as the noise variance $\tau^2$ increases. The SLM is found to be suitable only for clean data. Regarding the LSLM, the RMSE and bias are relatively small even when the noise variance $\tau^2$ is large; the LSLM is relatively robust against data noise.

**[Table 5 around here]**

Yet, the LSLM estimates tend to be biased when $N$ = 2,000 and $L$ is small. This result suggests that the LSLM faces a degeneracy problem, which is a well-known major problem of low rank modeling (Stein, 2014). Specifically, the absence of non-principal eigen-pairs can introduce errors into coefficients estimates. In our setting, this problem disappears when $L \geq 200$.

Table 6 summarizes the standard error of $\hat{\beta}_1$, which we denote se[$\hat{\beta}_1$]. Again, the LSLM and SLM tend to have smaller RMSEs and biases than the LM. However, the RMSEs and biases for the SLM are close to those for the LM estimates when data are noisy. The LSLM has small RMSEs and biases across cases, especially when $L \geq 200$. This outcome implies that consideration of data noise stabilizes the standard error estimates.

**[Table 6 around here]**

Table 7 summarizes results for the estimated $\hat{\rho}$ parameters. Interestingly, the LSLM estimates are nearly unbiased across cases, regardless of the number of eigen-pairs. The RMSEs are also quite small, especially when the sample size is large ($N$ = 2,000). By contrast, the standard SLM severely underestimates the $\rho$ value when data are not clean ($\tau^2 \neq 0$). Thus, although independent data noise has been ignored in almost all spatial econometric studies, this omission leads to an erroneous conclusion.

**[Table 7 around here]**

Tables 8 and 9 show RMSEs and biases for the estimated $DE_1$ and $IE_1$. $DE_k$ and $IE_k$ estimated for the LSLM have small RMSEs and biases across cases. Interestingly, any degeneracy problem due to the rank reduction is not detectable with this table; the increase in $N$ successfully improves estimation accuracy. Thus, while a low rank approach has never been used to estimate DE and IE, it accurately estimates these effects from large samples without encountering the degeneracy problem. The SLM again underestimates these effects when $\tau^2 \neq 0$. The underestimation for $IE_k$ is severe, especially when $\tau^2 = 4$ and $\rho = 0.8$.

**[Table 8 around here]**

**[Table 9 around here]**

Table 10 compares the mean lower and upper bounds of the 95% confidential intervals for $DE_1$ and $IE_1$ when $\rho = 0.6$ and $\tau^2 = 0.0$. The bootstrap approach is employed for the LSLM case, whereas the Markov chain Monte Carlo (MCMC) technique implemented in the R package spdep is used for the SLM. This table shows that these two approaches provide similar confidence intervals for $DE_1$. In contrast, our bootstrapping tends to have wider confidence intervals for $IE_1$ than for the SLM. This outcome might be because our approach estimates the spatial dependence parameters in each iteration, whereas the MCMC implementation in spdep assumes the spatial dependence parameter as given. Our result might be more reliable because it takes into account the uncertainty in these parameters.

[Table 10 around here]

### 4.2. Simulation experiment 2

This section compares the LM, LSEM, and SEM while assuming SEM to be the true data generating process. Section 4.2.1 summarizes our simulation experimental design, and Section 4.2.2 explains the simulation result.

### 4.2.1. The setting

Next we compare the LSEM and the standard SEM. The true data are generated from

$$\mathbf{y} = \beta_0 \mathbf{1} + \mathbf{z} + \mathbf{u} \qquad \mathbf{u} \sim N(\mathbf{0}, \tau^2 \mathbf{I})$$

$$\mathbf{z} = \beta_1 \mathbf{x}_1 + \beta_2 \mathbf{x}_2 + (\mathbf{I} - \varphi \mathbf{W})^{-1} \boldsymbol{\varepsilon}. \qquad \boldsymbol{\varepsilon} \sim N(\mathbf{0}, \mathbf{I})$$

(37)

where coefficients $\beta_0 = 1.0$, $\beta_1 = 2.0$, and $\beta_2 = 0.5$, which are the same setting used for the previous section. Covariates $\mathbf{x}_1$ and $\mathbf{x}_2$, and $\mathbf{W}$ are also given in the same way as for Sections 4.1 and 4.2. Eq. (37) coincides with the SEM when $\tau^2 = 0$. Simulations are replicated 200 times in each case, with $N \in \{500, 2000\}$, $\varphi \in \{0.2, 0.4, 0.6, 0.8\}$, and $\tau^2 \in \{0, 1, 2\}$.

The models being compared are as follows: the LM, LSEM$_{50}$, LSEM$_{100}$, LSEM$_{200}$, LSEM$_{300}$, LSEM$_{400}$, and SEM, where LSEM$_L$ represents LSEM with $L$ eigen-pairs.

### 4.2.2. Results

Table 11 summarizes RMSEs and biases of the estimated $\hat{\beta}_1$s. Estimates for all models are nearly unbiased and have small errors. This result is reasonable because both the

LM and SEM estimators are unbiased in the presence of spatially dependent errors.

**[Table 11 around here]**

Table 12 summarizes the estimated coefficients standard errors, se[$\hat{\beta}_1$]s. The RMSEs and biases for the LM increase when $\varphi$ is large and $\tau^2$ is small. Consideration of residual spatial dependence is needed to evaluate the coefficients standard errors appropriately. The coefficients standard errors, se[$\hat{\beta}_1$]s, estimated for the SEM are accurate across cases. The LSEM yields accurate estimates of se[$\hat{\beta}_1$]s, too. When $N$ = 2,000, the accuracy improves as the number of eigen-pairs $L$ increases. This outcome is because a larger number of eigen-pairs captures more spatial variation, and mitigates the degeneracy problem. In contrast, in many cases with $N$ = 500, the RMSEs are the smallest when $L$ = 200 or 300, but not when L = 400. This finding suggests that $L$ must be chosen to avoid overfitting when $N$ is small.

**[Table 12 around here]**

The left side of Table 13 summarizes the bias of the estimated $\hat{\varphi}$s. The $\hat{\varphi}$s estimated for the LSEM are biased when their true value is near zero. This outcome is because the LSEM employs not only the scale parameter $\varphi$, but also the variance parameter $\sigma^2$, to model residual spatial dependence. As a result, $\varphi = 0$ in the LSEM no longer means an absence of spatial dependence, whereas $\varphi = 0$ in the SEM does mean this. When using the LSEM, $\varphi$ must be interpreted as a pure scale parameter. Still, $\hat{\varphi}$ estimated for the LSEM indicates a small bias when $N$ is large and the true $\varphi$ is not small.

**[Table 13 around here]**

The right side of Table 13 summarizes the $z$-value of the Moran coefficient (see Cliff and Ord, 1981) for residual spatial dependence. This table verifies that the LSEM successfully reduces residual spatial dependence across cases. Nevertheless, when $N = 2,000$, the residual Moran coefficient tends to increase as $L$ decreases due to the degeneracy problem. Overcoming this problem would be an important next step.

### 4.3. A comparison of computational times

This section compares average computational times for the eigen-decomposition and the parameter estimation for the LSLM$_L$ and LSEM$_L$, where $L \in \{50, 100, 200\}$, using larger samples with $N \in \{5,000, 10,000, 20,000, 40,000\}$. The rARPACK package was used for the eigen-decomposition. The LSLM$_L$ was fitted 30 times to data generated with Eq. (34), and the LSEM$_L$ was fitted to those data generated with Eq. (37). The cost of ML estimation for the SLM and SEM is $O(N^3)$; roughly speaking, this ML estimation is infeasible when $N > 10,000$ in a standard computing environment.

Table 14 summarizes average computation times. The eigen-decomposition took only 148.1 seconds even when $N = 40,000$ and $L = 200$. Both of the LSLM and LSEM estimations require only seconds. These results confirm computational efficiency of our approach. Furthermore, despite bootstrapping all of the parameter estimates for the LSLM 200 times, the computationl time was only 176.9 seconds, even when $N = 40,000$ and $L = 200$, without parallelization. Our approach is especially suitable for large samples.

**[Table 14 around here]**

## 5. Concluding Remarks

This paper summarizes results of a study addressing the low rank spatial econometric modeling approach. The study's main features are as follow: it approximates standard spatial econometric models; it is consistent with a hierarchical model specification in geostatistics; it is computationally efficient; and, it accommodates data noise. Performance of the LSLM and SEM, the two basic specifications, are studied with Monte Carlo experiments. Results confirm estimation accuracy and computational efficiency.

Except for Tiefelsdorf and Griffith (2007), Griffith and Chun (2014), Chun et al. (2016), and some recent ESF studies, studies on a low rank approximation for SLM is still quite limited. Despite that, our results show that the LSLM accurately approximates the SLM. Especially, the *DE* and *IE* estimated for the LSLM have very small errors across cases. Rank reduction might be a sensible way to approximate the SLM.

Many issues remain to be addressed. First, our approaches must be applied to a wide variety of empirical case studies to clarify their advantages and disadvantages in practice.

Therein, sensitivity analyses of our approach with regard to the specification of matrix **W**, data distribution type, and so on, also is needed. Second, we need to mitigate the degeneracy problem due to rank reduction. Fortunately, recently low rank approaches mitigating this problem have been proposed in geostatistics (e.g., Sang and Huang, 2012, Nychka et al., 2015). Combining these approaches with our approach is an important next step to estimate coefficients accurately, even if the number of eigen-pairs is substantially smaller than $N$. Third, our approach must be extended to a wide variety of modeling problems. Fortunately, because our models are identical to the standard LME model (see Section 3.6), they could be extended to additive, multilevel, varying coefficients, and other models that have a LME model representation (Hodges, 2016). Besides, because our specification is identical to the hierarchical specification in geostatistics, potentially our models can be extended to dynamic spatiotemporal (e.g., Cressie and Johanessen, 2008), multivariate (e.g., Banerjee et al., 2008), and other models. These extensions would be useful for integrating spatial econometrics and geostatistics, which have been developed nearly independently.

Functions to implement the LSLM and LSEM are available in an R package spmoran (https://cran.r-project.org/web/packages/spmoran/index.html).

# Acknowledgement

This study was funded by the JSPS KAKENHI Grant Numbers 17K12974, 17K14738, and 18H03628.

# References

- Anselin, L. (1988) *Spatial Econometrics: Methods and Models*. Springer Science & Business Media, Berlin.

- Anselin, L. (2010) Thirty years of spatial econometrics. *Papers in regional science*, 89 (1), 3-25.

- Arbia, G. (2006) *Spatial Econometrics: statistical foundations and applications to regional convergence*. Springer, Berlin.

- Arbia, G., Espa, G., and Giuliani, D. (2016) Dirty spatial econometrics. *The Annals of Regional Science*, 56 (1), 177-189.

- Autant-Bernard, C. and LeSage, J. P. (2011) Quantifying knowledge spillovers using spatial econometric models. *Journal of regional Science*, 51 (3), 471-496.

- Banerjee, S., Gelfand, A.E., Finley, A.O., and Sang, H. (2008) Gaussian predictive process models for large spatial data sets. *Journal of the Royal Statistical Society: Series B (Statistical Methodology)*, 70 (4), 825-848.

- Banerjee, S., Carlin, B.P., and Gelfand, A.E. (2014) *Hierarchical modeling and analysis for spatial data*. CRC press, New York.

- Bates, D.M. (2010) *lme4: Mixed-Effects Modeling with R*. URL: http://lme4.r-forge.r-project.org/book/.

- Berlinear, L.M. (1996) Hierarchical Bayesian time-series models. In: Erickson, G. and Smith, C.R. (eds.) *Maximum Entropy and Bayesian Methods*, pp.15-22, Kluwer Academic Publishers, Dordrecht, Netherland.

- Burden, S., Cressie, N., and Steel, D.G. (2015) The SAR model for very large datasets: a reduced rank approach. *Econometrics*, 3 (2), 317-338.

- Cliff, A.D. and Ord, J. K. (1981) *Spatial Processes: Methods and Applications*. Pion, London, UK.

- Chun, Y., Griffith, D. A., Lee, M., and Sinha, P. (2016) Eigenvector selection with stepwise regression techniques to construct eigenvector spatial filters. *Journal of Geographical Systems*, 18 (1), 67-85.

- Cressie, N. (1993) *Statistics for Spatial Data*. Wiley, New York.

- Cressie, N. and Johannesson, G. (2008) Fixed rank kriging for very large spatial data sets. *Journal of the Royal Statistical Society: Series B (Statistical Methodology)*, 70 (1), 209-226.

- Cressie, N. and Wikle, C.K. (2011) *Statistics for Spatio-Temporal Data*. Wiley, Hoboken, New Jersey.

- Dall'Erba, S. and Le Gallo, J. (2008) Regional convergence and the impact of European structural funds over 1989–1999: A spatial econometric analysis. *Papers in Regional Science*, 87 (2), 219-244.

- Drineas, P. and Mahoney, M.W. (2005) On the Nyström method for approximating a Gram matrix for improved kernel-based learning. *Journal of Machine Learning Research*, 2153-2175.

- Elhorst, J.P. (2010) Applied spatial econometrics: raising the bar. *Spatial Economic Analysis*, 5 (1), 9-28.

- Elhorst, J.P. (2014) Linear spatial dependence models for cross-section data. In: *Spatial Econometrics: From Cross-sectional Data to Spatial Panels* (pp. 5-36). Springer, Berlin, Heidelberg.

- Gelfand, A.E., Diggle, P., Guttorp, P. and Fuentes, M. (2010) *Handbook of spatial statistics*. CRC press, New York.

- Gibbons, S., and Overman, H. G. (2012). Mostly pointless spatial econometrics?. Journal

of Regional Science, 52(2), 172-191.

- Griffith, D.A. (2000) Eigenfunction properties and approximations of selected incidence matrices employed in spatial analyses. *Linear Algebra and its Applications*, 321 (1-3), 95-112.

- Griffith, D.A. (2003) Spatial autocorrelation and spatial filtering: gaining understanding through theory and scientific visualization. *Springer Science & Business Media*, Berlin.

- Griffith, D.A. (2004) Faster maximum likelihood estimation of very large spatial autoregressive models: An extension of the Smirnov-Anselin result. *Journal of Statistical Computation and Simulation*, 74(12), 855-866.

- Griffith, D. A. (2015). Approximation of Gaussian spatial autoregressive models for massive regular square tessellation data. International Journal of Geographical Information Science, 29(12), 2143-2173.

- Griffith, D.A., Chun, Y. (2014) Spatial autocorrelation and spatial filtering. In: Fischer, M.M. and Nijkamp, N (eds). *Handbook of Regional Science*. Springer, Berlin, Heidelberg, pp. 1477–1507.

- Hodges, J. S. (2016). *Richly Parameterized Linear Models: Additive, Time Series, and*

*Spatial Models using Random Effects*. Chapman and Hall/CRC, New York.

- Kang, E. L. and Cressie, N. (2011). Bayesian inference for the spatial random effects model. *Journal of the American Statistical Association*, 106 (495), 972-983.

- Kelejian, H. and Piras, G. (2017) *Spatial Econometrics*, Academic Press, Elsevier, London.

- Kelejian, H.H. and Prucha, I.R. (1998) A generalized spatial two-stage least squares procedure for estimating a spatial autoregressive model with autoregressive disturbances. *The Journal of Real Estate Finance and Economics*, 17(1), 99-121.

- Kim, C. W., Phipps, T. T., and Anselin, L. (2003). Measuring the benefits of air quality improvement: a spatial hedonic approach. *Journal of Environmental Economics and Management*, 45 (1), 24-39.

- Lee, L.F. (2007) GMM and 2SLS estimation of mixed regressive, spatial autoregressive models. *Journal of Econometrics*, 137(2), 489-514.

- LeSage, J,P. and Pace, R.K, (2009) *Introduction to Spatial Econometrics*. CRC Press, New York.

- LeSage, J,P. and Pace, R.K, (2014) The biggest myth in spatial econometrics.

*Econometrics*, 2, 217-249.

- Manski, C.F. (1993) Identification of endogenous social effects: The reflection problem. *The Review of Economic Studies*, 60(3), 531-542.

- Murakami, D. and Griffith, D.A. (2015) Random effects specifications in eigenvector spatial filtering: a simulation study. *Journal of Geographical Systems,* 17 (4), 311-331.

- Murakami, D. and Griffith, D.A. (2018a) Eigenvector spatial filtering for large data sets: fixed and random effects approaches. *Geographical Analysis*. DOI: 10.1111/gean.12156.

- Murakami, D. and Griffith, D.A. (2018b) Spatially varying coefficient modeling for large datasets: Eliminating N from spatial regressions. *ArXiv*, 1807.09681.

- Murakami, D. and Seya, H. (2017) Spatially filtered unconditional quantile regression. *ArXiv*, 1706.07705.

- Nychka, D., Bandyopadhyay, S., Hammerling, D., Lindgren, F., and Sain, S. (2015) A multiresolution Gaussian process model for the analysis of large spatial datasets. *Journal of Computational and Graphical Statistics*, 24 (2), 579-599.

- Parent, O. and LeSage, J. P. (2008) Using the variance structure of the conditional autoregressive spatial specification to model knowledge spillovers. *Journal of Applied*

*Econometrics*, 23 (2), 235-256.

- Pinheiro, J.C. and Bates, D.M. (2000) *Mixed Effects Models in S and S-PLUS*. Springer, New York.

- Pratesi, M. and Salvati, N. (2008) Small area estimation: the EBLUP estimator based on spatially correlated random area effects. *Statistical Methods and Applications*, 17 (1), 113-141.

- Rey, S.J. and Montouri, B.D. (1999) US regional income convergence: a spatial econometric perspective. *Regional Studies*, 33 (2), 143-156.

- Salvati, N. (2004) Small area estimation by spatial models: the spatial empirical best linear unbiased prediction (spatial EBLUP). Working Paper, Department of Statistics, University of Florence.

- Sang, H. and Huang, J.Z. (2012) A full scale approximation of covariance functions for large spatial data sets. Journal of Royal Statistical Society: Series B (Statistical Methodology). 74 (1), 111-132.

- Seya, H., Tsutsumi, M., and Yamagata, Y. (2012) Income convergence in Japan: A Bayesian spatial Durbin model approach. *Economic Modelling*, 29 (1), 60-71.

- Seya, H., Yamagata, Y., and Tsutsumi, M. (2013) Automatic selection of a spatial weight matrix in spatial econometrics: Application to a spatial hedonic approach. *Regional Science and Urban Economics*, 43 (3), 429-444.

- Smirnov, O. and Anselin, L. (2001) Fast maximum likelihood estimation of very large spatial autoregressive models: a characteristic polynomial approach. *Computational Statistics and Data Analysis*, 35(3), 301-319.

- Tiefelsdorf, M. and Griffith, D. A. (2007) Semiparametric filtering of spatial autocorrelation: the eigenvector approach. *Environment and Planning A*, 39 (5), 1193-1221.

- Tsutsumi, M. and Seya, H. (2009) Hedonic approaches based on spatial econometrics and spatial statistics: application to evaluation of project benefits. *Journal of Geographical Systems*, 11 (4), 357-380.

- Upton, G. and Fingleton, B. (1985) *Spatial Data Analysis by Example. Volume 1: Point Pattern and Quantitative Data*. John Wiley & Sons, London, UK.

# Appendix.1 Fast parameter estimation

The restricted log-likelihood for the LSLM is given by Eqs.(24), (25), and (26) with $m_{yy} = \mathbf{y}'\mathbf{y}$ and $\mathbf{m}_{Ey} = \mathbf{E}'\mathbf{y}$. The inner products $\{\mathbf{M}_{XX}, \mathbf{M}_{EX}, \mathbf{m}_{Xy}\}$ are given as follows.

$\mathbf{M}_{XX}$ is given as

$$\mathbf{M}_{XX} = \begin{bmatrix} m_{11} & \mathbf{m}_{1X_{-1}} \\ \mathbf{m}'_{1X_{-1}} & \mathbf{M}_{X_{-1}X_{-1}} \end{bmatrix}, \quad (A1)$$

where

$$\mathbf{m}_{1X_{-1}} = \mathbf{1}'[\mathbf{X}_{-1} + \rho \mathbf{E}\Lambda(\mathbf{I}_L - \rho\Lambda)^{-1}\mathbf{E}'\mathbf{X}_{-1}]$$

$$= \mathbf{m}_{1X(-1)} + \mathbf{m}_{1E}\Lambda(\mathbf{I}_L - \rho\Lambda)^{-1}\mathbf{M}_{EX(-1)},$$

$$\mathbf{M}_{X_{-1}X_{-1}} = [\mathbf{X}_{-1} + \rho\mathbf{E}\Lambda(\mathbf{I}_L - \rho\Lambda)^{-1}\mathbf{E}'\mathbf{X}_{-1}]'[\mathbf{X}_{-1} + \rho\mathbf{E}\Lambda(\mathbf{I}_L - \rho\Lambda)^{-1}\mathbf{E}'\mathbf{X}_{-1}]$$

$$= \mathbf{M}_{XX(-1)} + 2\rho\mathbf{M}'_{EX(-1)}\Lambda(\mathbf{I}_L - \rho\Lambda)^{-1}\mathbf{M}_{EX(-1)}$$

$$+ \rho^2 \mathbf{M}'_{EX(-1)}\Lambda^2(\mathbf{I}_L - \rho\Lambda)^{-2}\mathbf{M}_{EX(-1)},$$

$m_{11} = \mathbf{1}'\mathbf{1} = N$, $\mathbf{m}_{1X(-1)} = \mathbf{1}'\mathbf{X}_{-1}$, $\mathbf{m}_{1E} = \mathbf{1}'\mathbf{E}$, $\mathbf{M}_{EX(-1)} = \mathbf{E}'\mathbf{X}_{-1}$, and $\mathbf{M}_{XX(-1)} = \mathbf{X}'_{-1}\mathbf{X}_{-1}$.

$\mathbf{M}_{EX}$ is given as

$$\mathbf{M}_{EX} = [\mathbf{m}'_{1E} \quad \mathbf{M}_{EX_{-1}}], \quad (A2)$$

where

$$\mathbf{M}_{EX_{-1}} = \mathbf{E}'(\mathbf{X}_{-1} + \rho\mathbf{E}\boldsymbol{\Lambda}(\mathbf{I}_L - \rho\boldsymbol{\Lambda})^{-1}\mathbf{E}'\mathbf{X}_{-1})$$

$$= \mathbf{M}_{EX(-1)} + \rho(\mathbf{I}_L - \rho\boldsymbol{\Lambda})^{-1}\mathbf{M}_{EX(-1)}.$$

$\mathbf{m}_{Xy}$ is given as

$$\mathbf{m}_{Xy} = [m_{1y} \quad \mathbf{m}'_{X(-1)y}]', \tag{A3}$$

where $m_{1y} = \mathbf{1}'\mathbf{y}$ and $\mathbf{m}_{X(-1)y} = \mathbf{X}'_{-1}\mathbf{y}$.

Once $\{m_{yy}, \mathbf{m}_{Ey}, m_{11}, \mathbf{m}_{1X(-1)}, \mathbf{m}_{1E}, \mathbf{M}_{EX(-1)}, \mathbf{M}_{XX(-1)}, m_{1y}, \mathbf{m}_{X(-1)y}\}$ are evaluated, Eqs. (A1) – (A3) no longer include matrix/vector whose size depends on $N$. Thus, the LSLM can be estimated computationally efficiently by evaluating these inner products first, and maximizing Eq.(24) after that.

As summarized in Table 4, the LSLM and the LSAC model are identical except that $\boldsymbol{\Sigma}_\theta$. Because the estimation procedure is unchanged even if $\boldsymbol{\Sigma}_\theta$ is changed, the fast estimation approach for the LSLM is readily available to the LSAC model. Likewise, the same procedure is applicable to the LSDM if only $\mathbf{X}_{-1}$ is replaced with $[\mathbf{X}_{-1}, \mathbf{W}\mathbf{X}_{-1}]$.

In the bootstrapping for the LSLM, the LSDM, and the LSAC model, because these models have the LME model representation (see Table 4), the bootstrap approach for the LME model, which was applied to the LSEM in Section 3.4, is also applicable to these models. As explained in Section 3.4, $\mathbf{X}$ and $\mathbf{E}$ are fixed and only $\mathbf{y}$ is changed across iterations. Thus, given $\{m_{11},\ \mathbf{m}_{1X(-1)},\ \mathbf{m}_{1E},\ \mathbf{M}_{EX(-1)},\ \mathbf{M}_{XX(-1)}\}$ before the iterations, the only elements we need to calculate in each iteration are $\{m_{yy}, \mathbf{m}_{Ey}, m_{1y}, \mathbf{m}_{X(-1)y}\}$ whose total number of operations is only $(K + L + 1)\,N$. Thus, the bootstrapping for these models are computationally efficient as same as the LSEM.

Table 1: Elements in the log-likelihood (Eq.5). Those for SDM are the same with SLM whose **X** is replaced with [**X**, **WX**] and **β** with [**β′**, **γ′**]′.

|   | SEM | SLM/SDM | SAC |
|---|---|---|---|
| **θ** | $\varphi$ | $\rho$ | $\{\varphi, \rho\}$ |
| **r̂** | $\mathbf{y} - \mathbf{X}\hat{\boldsymbol{\beta}}$ | $\mathbf{y} - (\mathbf{I} - \rho\mathbf{W})^{-1}\mathbf{X}\hat{\boldsymbol{\beta}}$ | $\mathbf{y} - (\mathbf{I} - \rho\mathbf{W})^{-1}\mathbf{X}\hat{\boldsymbol{\beta}}$ |
| **Z** | $\mathbf{I} - \varphi\mathbf{W}$ | $\mathbf{I} - \rho\mathbf{W}$ | $(\mathbf{I} - \varphi\mathbf{W})(\mathbf{I} - \rho\mathbf{W})$ |
| **β̂** | $[\mathbf{X}'(\mathbf{Z}'\mathbf{Z})^{-1}\mathbf{X}]^{-1}\mathbf{X}'(\mathbf{Z}'\mathbf{Z})^{-1}\mathbf{y}$ | $[\mathbf{X}'\mathbf{X}]^{-1}\mathbf{Z}\mathbf{y}$ | $[\mathbf{X}'(\mathbf{Z}'\mathbf{Z})^{-1}\mathbf{X}]^{-1}\mathbf{X}'(\mathbf{Z}'\mathbf{Z})^{-1}(\mathbf{I} - \rho\mathbf{W})^{-1}\mathbf{y}$ |

Table 2: Direct and indirect effects. $\hat{\beta}_k$ is the $k$-th element of $\hat{\boldsymbol{\beta}}$, and $q_k$ is the $k$-th element of $\hat{\mathbf{q}}$.

| | Definition | Average DE and IE | | |
|---|---|---|---|---|
| | | SEM | SLM/SAC | SDM |
| Direct effects | $DE_k = \partial y_i / \partial x_{i,k}$ | $\hat{\beta}_k$ | Mean of the diagonal elements of $(\mathbf{I} - \rho\mathbf{W})^{-1}\hat{\beta}_k$ | Mean of the diagonal elements of $(\mathbf{I} - \rho\mathbf{W})^{-1}(\hat{\beta}_k\mathbf{I} + \hat{q}_k\mathbf{W})$ |
| Indirect effects | $IE_k = \partial y_j / \partial x_{i,k}$ | 0 | Mean of the row-sums of the off-diagonal elements of $(\mathbf{I} - \rho\mathbf{W})^{-1}\hat{\beta}_k$ | Mean of the row-sums of the off-diagonal elements of $(\mathbf{I} - \rho\mathbf{W})^{-1}(\hat{\beta}_k\mathbf{I} + \hat{q}_k\mathbf{W})$ |

Table 3: Hierarchical representations of the low rank spatial econometric models

| | Data model | Process model | |
|---|---|---|---|
| LSEM | $\mathbf{y} = \beta_1 \mathbf{1} + \mathbf{z} + \mathbf{u}$, $\mathbf{u} \sim N(\mathbf{0}, \tau^2 \mathbf{I})$ | $\mathbf{z} = \mathbf{X}_{-1}\boldsymbol{\beta}_{-1} + \mathbf{E}\boldsymbol{\gamma}$ | $\boldsymbol{\gamma} \sim N(\mathbf{0}, \sigma^2(\mathbf{I} - \varphi\boldsymbol{\Lambda})^{-2})$ |
| LSLM | | $\mathbf{z} = [\mathbf{I} + \rho\mathbf{E}\boldsymbol{\Lambda}(\mathbf{I} - \rho\boldsymbol{\Lambda})^{-1}\mathbf{E}']\mathbf{X}_{-1}\boldsymbol{\beta}_{-1} + \mathbf{E}\boldsymbol{\gamma}$ | $\boldsymbol{\gamma} \sim N(\mathbf{0}, \sigma^2(\mathbf{I} - \rho\boldsymbol{\Lambda})^{-2})$ |
| LSAC | | | $\boldsymbol{\gamma} \sim N(\mathbf{0}, \sigma^2(\mathbf{I} - \rho\boldsymbol{\Lambda})^{-2}(\mathbf{I} - \varphi\boldsymbol{\Lambda})^{-2})$ |
| LSDM | | $\mathbf{z} = [\mathbf{I} + \rho\mathbf{E}\boldsymbol{\Lambda}(\mathbf{I} - \rho\boldsymbol{\Lambda})^{-1}\mathbf{E}'](\mathbf{X}_{-1}\boldsymbol{\beta}_{-1} + \mathbf{W}\mathbf{X}_{-1}\boldsymbol{\gamma}_{-1}) + \mathbf{E}\boldsymbol{\gamma}$ | $\boldsymbol{\gamma} \sim N(\mathbf{0}, \sigma^2(\mathbf{I} - \rho\boldsymbol{\Lambda})^{-2})$ |

Table 4: Variables/parameters in Eq.(23)

| | $\mathbf{X_\theta}$ | $\mathbf{\Sigma_\theta}$ |
|---|---|---|
| LSEM | $\mathbf{X}$ | $(\sigma/\tau)\,(\mathbf{I}-\varphi\mathbf{\Lambda})^{-1}$ |
| LSLM | $[\mathbf{1}, \{\mathbf{I}+\rho\mathbf{E}\mathbf{\Lambda}(\mathbf{I}-\rho\mathbf{\Lambda})^{-1}\mathbf{E}'\}\mathbf{X}_{-1}]$ | $(\sigma/\tau)\,(\mathbf{I}-\rho\mathbf{\Lambda})^{-1}$ |
| LSAC | | $(\sigma/\tau)\,(\mathbf{I}-\varphi\mathbf{\Lambda})^{-1}(\mathbf{I}-\rho\mathbf{\Lambda})^{-1}$ |
| LSDM | $[\mathbf{1}, \{\mathbf{I}+\rho\mathbf{E}\mathbf{\Lambda}(\mathbf{I}-\rho\mathbf{\Lambda})^{-1}\mathbf{E}'\}\mathbf{X}_{-1}, \{\mathbf{I}+\rho\mathbf{E}\mathbf{\Lambda}(\mathbf{I}-\rho\mathbf{\Lambda})^{-1}\mathbf{E}'\}\mathbf{W}\mathbf{X}_{-1}]$ | $(\sigma/\tau)\,(\mathbf{I}-\rho\mathbf{\Lambda})^{-1}$ |

Table 5: RMSE and Bias of $\beta_1$ (LM, LSLM, and SLM). Darker red represents larger RMSE and bias.

| N | $\rho$ | $\tau^2$ | RMSE | | | | | | | Bias | | | | | | |
|---|---|---|---|---|---|---|---|---|---|---|---|---|---|---|---|---|
| | | | LM | LSLM | | | | | SLM | LM | LSLM | | | | | SLM |
| | | | | 50 | 100 | 200 | 300 | 400 | | | 50 | 100 | 200 | 300 | 400 | |
| 500 | 0.2 | 0.0 | 0.05 | 0.06 | 0.07 | 0.05 | 0.05 | 0.05 | 0.04 | 0.01 | -0.03 | -0.05 | -0.03 | -0.02 | -0.01 | -0.01 |
| | | 2.0 | 0.11 | 0.11 | 0.11 | 0.11 | 0.11 | 0.11 | 0.11 | 0.02 | -0.03 | -0.04 | -0.02 | -0.01 | 0.00 | 0.01 |
| | | 4.0 | 0.18 | 0.20 | 0.20 | 0.19 | 0.19 | 0.19 | 0.18 | 0.02 | -0.02 | -0.03 | -0.02 | -0.01 | -0.01 | 0.01 |
| | 0.4 | 0.0 | 0.08 | 0.07 | 0.09 | 0.07 | 0.06 | 0.05 | 0.05 | 0.06 | -0.04 | -0.08 | -0.05 | -0.03 | -0.02 | -0.01 |
| | | 2.0 | 0.13 | 0.11 | 0.13 | 0.11 | 0.11 | 0.11 | 0.11 | 0.07 | -0.04 | -0.07 | -0.04 | -0.02 | -0.01 | 0.03 |
| | | 4.0 | 0.20 | 0.20 | 0.21 | 0.20 | 0.20 | 0.19 | 0.19 | 0.07 | -0.03 | -0.05 | -0.03 | -0.01 | 0.00 | 0.05 |
| | 0.6 | 0.0 | 0.19 | 0.07 | 0.11 | 0.08 | 0.06 | 0.06 | 0.04 | 0.18 | -0.04 | -0.10 | -0.07 | -0.04 | -0.03 | 0.00 |
| | | 2.0 | 0.22 | 0.11 | 0.13 | 0.12 | 0.11 | 0.11 | 0.13 | 0.18 | -0.02 | -0.08 | -0.05 | -0.02 | 0.00 | 0.07 |
| | | 4.0 | 0.26 | 0.21 | 0.22 | 0.21 | 0.21 | 0.21 | 0.23 | 0.16 | -0.04 | -0.09 | -0.06 | -0.03 | -0.01 | 0.10 |
| | 0.8 | 0.0 | 0.47 | 0.06 | 0.11 | 0.09 | 0.07 | 0.06 | 0.05 | 0.46 | -0.01 | -0.10 | -0.08 | -0.05 | -0.03 | 0.01 |
| | | 2.0 | 0.49 | 0.11 | 0.13 | 0.12 | 0.10 | 0.10 | 0.16 | 0.46 | 0.01 | -0.08 | -0.05 | -0.02 | 0.00 | 0.11 |
| | | 4.0 | 0.50 | 0.21 | 0.21 | 0.20 | 0.20 | 0.20 | 0.30 | 0.43 | -0.01 | -0.08 | -0.05 | -0.02 | -0.01 | 0.21 |
| 2000 | 0.2 | 0.0 | 0.02 | 0.02 | 0.02 | 0.04 | 0.05 | 0.05 | 0.02 | 0.01 | 0.00 | -0.01 | -0.03 | -0.05 | -0.04 | 0.00 |
| | | 2.0 | 0.05 | 0.05 | 0.05 | 0.06 | 0.07 | 0.07 | 0.05 | 0.01 | 0.00 | -0.01 | -0.03 | -0.05 | -0.04 | 0.00 |
| | | 4.0 | 0.09 | 0.09 | 0.09 | 0.10 | 0.10 | 0.10 | 0.09 | 0.03 | 0.02 | 0.00 | -0.01 | -0.03 | -0.03 | 0.02 |
| | 0.4 | 0.0 | 0.07 | 0.04 | 0.02 | 0.05 | 0.08 | 0.08 | 0.02 | 0.06 | 0.03 | 0.00 | -0.04 | -0.08 | -0.08 | 0.00 |
| | | 2.0 | 0.08 | 0.06 | 0.05 | 0.07 | 0.09 | 0.09 | 0.06 | 0.06 | 0.03 | 0.00 | -0.04 | -0.08 | -0.07 | 0.03 |
| | | 4.0 | 0.12 | 0.11 | 0.10 | 0.10 | 0.11 | 0.11 | 0.11 | 0.08 | 0.05 | 0.02 | -0.02 | -0.06 | -0.05 | 0.06 |
| | 0.6 | 0.0 | 0.18 | 0.11 | 0.06 | 0.05 | 0.10 | 0.10 | 0.02 | 0.18 | 0.11 | 0.05 | -0.04 | -0.09 | -0.09 | 0.00 |
| | | 2.0 | 0.19 | 0.12 | 0.08 | 0.07 | 0.11 | 0.11 | 0.09 | 0.17 | 0.11 | 0.05 | -0.04 | -0.09 | -0.09 | 0.06 |
| | | 4.0 | 0.21 | 0.15 | 0.12 | 0.10 | 0.13 | 0.12 | 0.16 | 0.19 | 0.12 | 0.06 | -0.02 | -0.08 | -0.08 | 0.13 |
| | 0.8 | 0.0 | 0.46 | 0.28 | 0.16 | 0.03 | 0.09 | 0.10 | 0.02 | 0.46 | 0.28 | 0.16 | 0.00 | -0.09 | -0.09 | 0.00 |
| | | 2.0 | 0.46 | 0.28 | 0.16 | 0.05 | 0.10 | 0.10 | 0.12 | 0.46 | 0.27 | 0.15 | 0.00 | -0.09 | -0.09 | 0.10 |
| | | 4.0 | 0.48 | 0.30 | 0.19 | 0.10 | 0.12 | 0.12 | 0.26 | 0.46 | 0.28 | 0.16 | 0.01 | -0.07 | -0.07 | 0.24 |

Table 6: RMSE and Bias of se[$\beta_1$] (LM, LSLM, and SLM). Darker red represents larger RMSE and bias.

| N | $\rho$ | $\tau^2$ | RMSE | | | | | | | Bias | | | | | | |
|---|---|---|---|---|---|---|---|---|---|---|---|---|---|---|---|---|
| | | | LM | LSLM | | | | | SLM | LM | LSLM | | | | | SLM |
| | | | | 50 | 100 | 200 | 300 | 400 | | | 50 | 100 | 200 | 300 | 400 | |
| 500 | 0.2 | 0.0 | 0.002 | 0.002 | 0.002 | 0.003 | 0.003 | 0.004 | 0.001 | 0.001 | -0.001 | -0.002 | -0.002 | -0.002 | -0.003 | 0.000 |
| | | 2.0 | 0.003 | 0.005 | 0.005 | 0.005 | 0.005 | 0.006 | 0.003 | 0.001 | -0.002 | -0.003 | -0.003 | -0.003 | -0.004 | 0.000 |
| | | 4.0 | 0.006 | 0.008 | 0.009 | 0.009 | 0.010 | 0.011 | 0.006 | 0.002 | -0.003 | -0.005 | -0.005 | -0.005 | -0.007 | 0.001 |
| | 0.4 | 0.0 | 0.006 | 0.002 | 0.003 | 0.004 | 0.005 | 0.007 | 0.002 | 0.006 | 0.000 | -0.003 | -0.004 | -0.005 | -0.007 | 0.000 |
| | | 2.0 | 0.006 | 0.005 | 0.007 | 0.006 | 0.006 | 0.007 | 0.005 | 0.005 | -0.003 | -0.005 | -0.005 | -0.005 | -0.006 | 0.003 |
| | | 4.0 | 0.011 | 0.009 | 0.011 | 0.011 | 0.010 | 0.011 | 0.010 | 0.009 | -0.003 | -0.006 | -0.006 | -0.005 | -0.006 | 0.008 |
| | 0.6 | 0.0 | 0.018 | 0.002 | 0.003 | 0.005 | 0.006 | 0.009 | 0.002 | 0.018 | 0.002 | -0.003 | -0.004 | -0.006 | -0.009 | 0.001 |
| | | 2.0 | 0.016 | 0.004 | 0.007 | 0.007 | 0.008 | 0.009 | 0.009 | 0.015 | -0.002 | -0.006 | -0.007 | -0.007 | -0.008 | 0.009 |
| | | 4.0 | 0.025 | 0.011 | 0.013 | 0.012 | 0.011 | 0.011 | 0.023 | 0.024 | -0.002 | -0.008 | -0.007 | -0.006 | -0.007 | 0.021 |
| | 0.8 | 0.0 | 0.056 | 0.006 | 0.002 | 0.005 | 0.007 | 0.010 | 0.002 | 0.056 | 0.006 | -0.001 | -0.004 | -0.007 | -0.010 | 0.002 |
| | | 2.0 | 0.045 | 0.004 | 0.008 | 0.008 | 0.009 | 0.010 | 0.017 | 0.045 | 0.000 | -0.007 | -0.008 | -0.009 | -0.010 | 0.017 |
| | | 4.0 | 0.059 | 0.012 | 0.014 | 0.013 | 0.012 | 0.012 | 0.047 | 0.059 | 0.000 | -0.009 | -0.008 | -0.008 | -0.008 | 0.046 |
| 2000 | 0.2 | 0.0 | 0.001 | 0.000 | 0.000 | 0.001 | 0.001 | 0.001 | 0.000 | 0.001 | 0.000 | 0.000 | 0.000 | -0.001 | -0.001 | 0.000 |
| | | 2.0 | 0.001 | 0.001 | 0.001 | 0.001 | 0.002 | 0.002 | 0.001 | 0.001 | 0.000 | 0.000 | -0.001 | -0.001 | -0.001 | 0.000 |
| | | 4.0 | 0.002 | 0.002 | 0.002 | 0.002 | 0.003 | 0.003 | 0.002 | 0.001 | 0.000 | 0.000 | -0.001 | -0.002 | -0.002 | 0.001 |
| | 0.4 | 0.0 | 0.003 | 0.002 | 0.001 | 0.000 | 0.001 | 0.001 | 0.000 | 0.003 | 0.002 | 0.001 | 0.000 | -0.001 | -0.001 | 0.000 |
| | | 2.0 | 0.003 | 0.002 | 0.001 | 0.002 | 0.002 | 0.003 | 0.002 | 0.003 | 0.001 | 0.000 | -0.001 | -0.002 | -0.002 | 0.002 |
| | | 4.0 | 0.005 | 0.003 | 0.002 | 0.003 | 0.004 | 0.004 | 0.005 | 0.005 | 0.003 | 0.001 | -0.001 | -0.003 | -0.003 | 0.004 |
| | 0.6 | 0.0 | 0.009 | 0.006 | 0.004 | 0.001 | 0.001 | 0.001 | 0.001 | 0.009 | 0.006 | 0.004 | 0.001 | -0.001 | -0.001 | 0.001 |
| | | 2.0 | 0.008 | 0.005 | 0.003 | 0.001 | 0.003 | 0.003 | 0.005 | 0.008 | 0.005 | 0.002 | -0.001 | -0.003 | -0.003 | 0.005 |
| | | 4.0 | 0.013 | 0.008 | 0.005 | 0.003 | 0.004 | 0.004 | 0.012 | 0.013 | 0.008 | 0.004 | -0.001 | -0.003 | -0.004 | 0.012 |
| | 0.8 | 0.0 | 0.028 | 0.016 | 0.010 | 0.003 | 0.000 | 0.001 | 0.001 | 0.028 | 0.016 | 0.010 | 0.003 | 0.000 | -0.001 | 0.001 |
| | | 2.0 | 0.023 | 0.012 | 0.006 | 0.001 | 0.003 | 0.003 | 0.009 | 0.023 | 0.012 | 0.006 | 0.000 | -0.003 | -0.003 | 0.008 |
| | | 4.0 | 0.030 | 0.017 | 0.009 | 0.003 | 0.004 | 0.004 | 0.024 | 0.030 | 0.016 | 0.009 | 0.001 | -0.003 | -0.004 | 0.024 |

Table 7: RMSE and Bias of $\rho$ (LSLM and SLM). Darker red represents larger RMSE and bias.

| N | $\rho$ | $\tau^2$ | RMSE LSLM 50 | 100 | 200 | 300 | 400 | SLM | Bias LSLM 50 | 100 | 200 | 300 | 400 | SLM |
|---|---|---|---|---|---|---|---|---|---|---|---|---|---|---|
| 500 | 0.2 | 0.0 | 0.09 | 0.08 | 0.07 | 0.06 | 0.06 | 0.04 | 0.01 | 0.03 | 0.03 | 0.03 | 0.03 | -0.01 |
| | | 2.0 | 0.17 | 0.15 | 0.13 | 0.12 | 0.11 | 0.13 | 0.01 | 0.02 | 0.02 | 0.02 | 0.01 | -0.11 |
| | | 4.0 | 0.33 | 0.29 | 0.23 | 0.22 | 0.22 | 0.19 | -0.07 | -0.05 | -0.02 | -0.02 | -0.02 | -0.17 |
| | 0.4 | 0.0 | 0.07 | 0.08 | 0.07 | 0.07 | 0.07 | 0.04 | 0.03 | 0.06 | 0.06 | 0.05 | 0.06 | -0.02 |
| | | 2.0 | 0.12 | 0.11 | 0.10 | 0.10 | 0.09 | 0.21 | 0.03 | 0.05 | 0.04 | 0.03 | 0.03 | -0.20 |
| | | 4.0 | 0.22 | 0.21 | 0.17 | 0.17 | 0.17 | 0.33 | -0.02 | -0.01 | 0.00 | 0.00 | -0.01 | -0.32 |
| | 0.6 | 0.0 | 0.07 | 0.08 | 0.08 | 0.08 | 0.08 | 0.03 | 0.05 | 0.07 | 0.07 | 0.07 | 0.07 | -0.02 |
| | | 2.0 | 0.07 | 0.08 | 0.07 | 0.07 | 0.06 | 0.24 | 0.03 | 0.05 | 0.05 | 0.04 | 0.04 | -0.23 |
| | | 4.0 | 0.12 | 0.11 | 0.10 | 0.10 | 0.10 | 0.43 | 0.02 | 0.04 | 0.03 | 0.02 | 0.02 | -0.42 |
| | 0.8 | 0.0 | 0.05 | 0.07 | 0.07 | 0.07 | 0.07 | 0.02 | 0.05 | 0.06 | 0.07 | 0.07 | 0.07 | -0.01 |
| | | 2.0 | 0.05 | 0.06 | 0.06 | 0.05 | 0.05 | 0.20 | 0.04 | 0.05 | 0.05 | 0.04 | 0.04 | -0.19 |
| | | 4.0 | 0.06 | 0.06 | 0.06 | 0.06 | 0.06 | 0.42 | 0.03 | 0.04 | 0.04 | 0.04 | 0.03 | -0.42 |
| 2000 | 0.2 | 0.0 | 0.06 | 0.04 | 0.04 | 0.05 | 0.05 | 0.02 | 0.00 | 0.01 | 0.02 | 0.03 | 0.03 | -0.01 |
| | | 2.0 | 0.12 | 0.09 | 0.07 | 0.07 | 0.07 | 0.11 | -0.01 | 0.00 | 0.02 | 0.03 | 0.03 | -0.11 |
| | | 4.0 | 0.27 | 0.19 | 0.15 | 0.13 | 0.13 | 0.17 | -0.07 | -0.02 | -0.01 | 0.01 | 0.01 | -0.16 |
| | 0.4 | 0.0 | 0.05 | 0.04 | 0.04 | 0.06 | 0.06 | 0.02 | 0.00 | 0.01 | 0.04 | 0.05 | 0.06 | -0.01 |
| | | 2.0 | 0.07 | 0.05 | 0.06 | 0.06 | 0.06 | 0.20 | 0.00 | 0.01 | 0.03 | 0.05 | 0.05 | -0.19 |
| | | 4.0 | 0.17 | 0.11 | 0.10 | 0.09 | 0.09 | 0.31 | -0.04 | 0.00 | 0.02 | 0.04 | 0.04 | -0.31 |
| | 0.6 | 0.0 | 0.03 | 0.03 | 0.05 | 0.06 | 0.07 | 0.02 | 0.01 | 0.02 | 0.04 | 0.06 | 0.07 | -0.01 |
| | | 2.0 | 0.04 | 0.04 | 0.05 | 0.06 | 0.06 | 0.23 | 0.00 | 0.02 | 0.04 | 0.06 | 0.06 | -0.23 |
| | | 4.0 | 0.07 | 0.06 | 0.06 | 0.07 | 0.07 | 0.42 | -0.01 | 0.01 | 0.03 | 0.05 | 0.05 | -0.42 |
| | 0.8 | 0.0 | 0.02 | 0.03 | 0.05 | 0.06 | 0.06 | 0.01 | 0.01 | 0.02 | 0.04 | 0.06 | 0.06 | -0.01 |
| | | 2.0 | 0.02 | 0.03 | 0.04 | 0.06 | 0.06 | 0.19 | 0.01 | 0.02 | 0.04 | 0.05 | 0.06 | -0.19 |
| | | 4.0 | 0.03 | 0.03 | 0.04 | 0.05 | 0.05 | 0.41 | 0.01 | 0.02 | 0.04 | 0.05 | 0.05 | -0.41 |

Table 8: RMSE and Bias of $DE_1$ (LSLM, and SLM). Darker red represents larger RMSE and bias.

| N | $\rho$ | $\tau^2$ | RMSE | | | | | | Bias | | | | | |
|---|---|---|---|---|---|---|---|---|---|---|---|---|---|---|
| | | | LSLM | | | | | SLM | LSLM | | | | | SLM |
| | | | 50 | 100 | 200 | 300 | 400 | | 50 | 100 | 200 | 300 | 400 | |
| 500 | 0.2 | 0.0 | 0.05 | 0.05 | 0.04 | 0.04 | 0.04 | 0.05 | -0.01 | -0.01 | -0.01 | -0.01 | -0.01 | -0.01 |
| | | 2.0 | 0.11 | 0.11 | 0.11 | 0.11 | 0.11 | 0.11 | 0.01 | 0.01 | 0.01 | 0.01 | 0.01 | 0.00 |
| | | 4.0 | 0.18 | 0.18 | 0.19 | 0.18 | 0.19 | 0.18 | 0.01 | 0.01 | 0.01 | 0.01 | 0.01 | 0.00 |
| | 0.4 | 0.0 | 0.05 | 0.05 | 0.05 | 0.05 | 0.05 | 0.05 | -0.01 | -0.01 | -0.01 | -0.01 | -0.01 | -0.01 |
| | | 2.0 | 0.11 | 0.11 | 0.11 | 0.11 | 0.11 | 0.11 | 0.01 | 0.00 | 0.00 | 0.00 | 0.00 | -0.02 |
| | | 4.0 | 0.18 | 0.19 | 0.19 | 0.18 | 0.19 | 0.19 | 0.01 | 0.01 | 0.01 | 0.01 | 0.01 | -0.01 |
| | 0.6 | 0.0 | 0.06 | 0.06 | 0.05 | 0.05 | 0.06 | 0.05 | 0.00 | 0.00 | 0.00 | 0.00 | 0.01 | -0.01 |
| | | 2.0 | 0.11 | 0.11 | 0.11 | 0.11 | 0.11 | 0.12 | 0.01 | 0.01 | 0.00 | 0.00 | 0.01 | -0.05 |
| | | 4.0 | 0.20 | 0.19 | 0.19 | 0.19 | 0.19 | 0.22 | -0.01 | -0.01 | -0.01 | -0.01 | -0.01 | -0.07 |
| | 0.8 | 0.0 | 0.08 | 0.08 | 0.07 | 0.08 | 0.08 | 0.07 | 0.01 | 0.01 | 0.02 | 0.03 | 0.04 | -0.01 |
| | | 2.0 | 0.12 | 0.12 | 0.11 | 0.11 | 0.11 | 0.20 | 0.00 | 0.01 | 0.00 | 0.00 | 0.00 | -0.16 |
| | | 4.0 | 0.22 | 0.21 | 0.20 | 0.20 | 0.20 | 0.30 | -0.02 | -0.01 | -0.02 | -0.02 | -0.02 | -0.20 |
| 2000 | 0.2 | 0.0 | 0.02 | 0.02 | 0.02 | 0.02 | 0.02 | 0.02 | 0.00 | 0.00 | 0.00 | 0.00 | 0.00 | 0.00 |
| | | 2.0 | 0.05 | 0.05 | 0.05 | 0.05 | 0.05 | 0.05 | 0.00 | 0.00 | 0.00 | 0.00 | -0.01 | -0.01 |
| | | 4.0 | 0.09 | 0.09 | 0.09 | 0.09 | 0.09 | 0.09 | 0.01 | 0.01 | 0.01 | 0.01 | 0.01 | 0.01 |
| | 0.4 | 0.0 | 0.03 | 0.03 | 0.02 | 0.02 | 0.02 | 0.02 | 0.00 | 0.00 | 0.00 | 0.00 | 0.00 | -0.01 |
| | | 2.0 | 0.05 | 0.05 | 0.05 | 0.05 | 0.05 | 0.06 | 0.00 | 0.00 | 0.00 | -0.01 | -0.01 | -0.02 |
| | | 4.0 | 0.09 | 0.09 | 0.09 | 0.09 | 0.09 | 0.09 | 0.01 | 0.01 | 0.01 | 0.01 | 0.01 | 0.00 |
| | 0.6 | 0.0 | 0.03 | 0.03 | 0.03 | 0.03 | 0.03 | 0.03 | 0.00 | 0.00 | 0.00 | 0.00 | 0.00 | -0.01 |
| | | 2.0 | 0.06 | 0.06 | 0.06 | 0.06 | 0.06 | 0.09 | -0.01 | -0.01 | -0.01 | -0.01 | -0.01 | -0.06 |
| | | 4.0 | 0.10 | 0.10 | 0.10 | 0.10 | 0.10 | 0.10 | 0.01 | 0.01 | 0.01 | 0.01 | 0.01 | -0.04 |
| | 0.8 | 0.0 | 0.05 | 0.04 | 0.04 | 0.03 | 0.03 | 0.03 | 0.00 | 0.00 | 0.00 | 0.00 | 0.01 | -0.02 |
| | | 2.0 | 0.07 | 0.06 | 0.06 | 0.06 | 0.06 | 0.17 | 0.00 | -0.01 | 0.00 | 0.00 | 0.00 | -0.16 |
| | | 4.0 | 0.10 | 0.10 | 0.10 | 0.10 | 0.10 | 0.19 | 0.00 | 0.00 | 0.00 | 0.01 | 0.01 | -0.15 |

Table 9: RMSE and Bias of $IE_1$ (LSLM, and SLM). Darker red represents larger RMSE and bias.

| N | $\rho$ | $\tau^2$ | RMSE LSLM 50 | 100 | 200 | 300 | 400 | SLM | Bias LSLM 50 | 100 | 200 | 300 | 400 | SLM |
|---|---|---|---|---|---|---|---|---|---|---|---|---|---|---|
| 500 | 0.2 | 0.0 | 0.22 | 0.19 | 0.18 | 0.18 | 0.18 | 0.13 | -0.01 | 0.04 | 0.04 | 0.05 | 0.06 | -0.02 |
|  |  | 2.0 | 0.44 | 0.39 | 0.35 | 0.32 | 0.32 | 0.33 | 0.06 | 0.06 | 0.05 | 0.05 | 0.05 | -0.29 |
|  |  | 4.0 | 0.73 | 0.67 | 0.63 | 0.60 | 0.59 | 0.46 | 0.00 | -0.02 | 0.05 | 0.04 | 0.03 | -0.42 |
|  | 0.4 | 0.0 | 0.32 | 0.30 | 0.31 | 0.31 | 0.32 | 0.21 | -0.01 | 0.10 | 0.15 | 0.17 | 0.19 | -0.07 |
|  |  | 2.0 | 0.56 | 0.53 | 0.49 | 0.47 | 0.45 | 0.81 | 0.08 | 0.11 | 0.11 | 0.10 | 0.09 | -0.78 |
|  |  | 4.0 | 0.87 | 0.85 | 0.81 | 0.77 | 0.76 | 1.14 | 0.01 | 0.01 | 0.05 | 0.03 | 0.03 | -1.12 |
|  | 0.6 | 0.0 | 0.64 | 0.69 | 0.75 | 0.76 | 0.76 | 0.39 | 0.16 | 0.37 | 0.50 | 0.55 | 0.57 | -0.17 |
|  |  | 2.0 | 0.73 | 0.72 | 0.70 | 0.68 | 0.66 | 1.86 | 0.02 | 0.16 | 0.19 | 0.17 | 0.15 | -1.84 |
|  |  | 4.0 | 1.27 | 1.21 | 1.16 | 1.12 | 1.09 | 2.59 | 0.04 | 0.17 | 0.13 | 0.09 | 0.10 | -2.58 |
|  | 0.8 | 0.0 | 1.84 | 2.33 | 2.75 | 2.87 | 2.87 | 1.31 | 0.58 | 1.48 | 2.03 | 2.20 | 2.22 | -0.51 |
|  |  | 2.0 | 1.86 | 2.00 | 1.95 | 1.86 | 1.77 | 6.03 | -0.04 | 0.49 | 0.49 | 0.39 | 0.26 | -6.00 |
|  |  | 4.0 | 2.71 | 2.76 | 2.70 | 2.65 | 2.60 | 8.00 | -0.08 | 0.30 | 0.23 | 0.10 | 0.04 | -7.99 |
| 2000 | 0.2 | 0.0 | 0.17 | 0.12 | 0.10 | 0.10 | 0.10 | 0.07 | -0.02 | -0.02 | 0.00 | 0.02 | 0.03 | -0.02 |
|  |  | 2.0 | 0.31 | 0.23 | 0.18 | 0.18 | 0.17 | 0.30 | -0.03 | -0.01 | 0.00 | 0.02 | 0.01 | -0.29 |
|  |  | 4.0 | 0.58 | 0.46 | 0.38 | 0.34 | 0.33 | 0.42 | -0.08 | 0.00 | -0.02 | 0.01 | 0.01 | -0.42 |
|  | 0.4 | 0.0 | 0.24 | 0.18 | 0.14 | 0.16 | 0.17 | 0.12 | -0.04 | -0.03 | 0.01 | 0.06 | 0.09 | -0.07 |
|  |  | 2.0 | 0.35 | 0.26 | 0.22 | 0.23 | 0.22 | 0.79 | -0.04 | -0.02 | 0.01 | 0.05 | 0.05 | -0.79 |
|  |  | 4.0 | 0.65 | 0.51 | 0.44 | 0.41 | 0.41 | 1.11 | -0.10 | -0.01 | -0.01 | 0.04 | 0.04 | -1.11 |
|  | 0.6 | 0.0 | 0.37 | 0.29 | 0.25 | 0.31 | 0.36 | 0.24 | -0.07 | -0.05 | 0.05 | 0.18 | 0.27 | -0.16 |
|  |  | 2.0 | 0.47 | 0.37 | 0.34 | 0.36 | 0.37 | 1.86 | -0.09 | -0.07 | 0.01 | 0.12 | 0.14 | -1.86 |
|  |  | 4.0 | 0.77 | 0.64 | 0.60 | 0.60 | 0.60 | 2.58 | -0.13 | -0.01 | 0.03 | 0.12 | 0.13 | -2.57 |
|  | 0.8 | 0.0 | 1.10 | 0.91 | 0.75 | 1.03 | 1.25 | 0.79 | -0.39 | -0.38 | 0.12 | 0.70 | 0.99 | -0.56 |
|  |  | 2.0 | 1.19 | 0.98 | 0.90 | 1.02 | 1.05 | 5.95 | -0.42 | -0.39 | 0.06 | 0.48 | 0.55 | -5.94 |
|  |  | 4.0 | 1.47 | 1.27 | 1.23 | 1.27 | 1.28 | 7.97 | -0.62 | -0.51 | -0.12 | 0.18 | 0.21 | -7.97 |

Table 10: Mean lower and upper bounds of the estimated 95 % confidence intervals for $DE_1$ and $IE_1$ ($\rho = 0.6$; $\tau^2 = 0.0$). The bootstrapping with 200 iterations is used for LSLM and the MCMC implemented in the impact function is an R package spdep is used for SLM.

|  | N | Lower bound | | | | Upper bound | | | |
|---|---|---|---|---|---|---|---|---|---|
|  |  | LSLM | | | SLM | LSLM | | | SLM |
|  |  | 50 | 100 | 200 |  | 50 | 100 | 200 |  |
| $DE_1$ | 500 | 2.20 | 2.21 | 2.21 | 2.22 | 2.46 | 2.45 | 2.44 | 2.43 |
|  | 2,000 | 2.25 | 2.25 | 2.26 | 2.27 | 2.39 | 2.38 | 2.39 | 2.38 |
| $IE_1$ | 500 | 1.78 | 2.13 | 2.47 | 2.26 | 3.78 | 4.02 | 4.11 | 3.09 |
|  | 2,000 | 1.63 | 1.98 | 2.32 | 2.45 | 3.04 | 3.11 | 3.31 | 2.86 |

Table 11: RMSE and Bias of $\beta_1$ ((LM, LSEM, and SEM). Darker red represents larger RMSE and bias.

| N | $\varphi$ | $\tau^2$ | RMSE | | | | | | | Bias | | | | | | |
|---|---|---|---|---|---|---|---|---|---|---|---|---|---|---|---|---|
| | | | LM | LSEM | | | | | SEM | LM | LSEM | | | | | SEM |
| | | | | 50 | 100 | 200 | 300 | 400 | | | 50 | 100 | 200 | 300 | 400 | |
| 500 | 0.2 | 0.0 | 0.05 | 0.05 | 0.05 | 0.05 | 0.05 | 0.05 | 0.05 | 0.00 | 0.00 | 0.00 | 0.00 | 0.00 | 0.00 | 0.00 |
| | | 2.0 | 0.06 | 0.06 | 0.06 | 0.06 | 0.06 | 0.06 | 0.06 | 0.00 | 0.00 | 0.00 | 0.00 | 0.00 | 0.00 | 0.00 |
| | | 4.0 | 0.10 | 0.10 | 0.10 | 0.10 | 0.10 | 0.10 | 0.10 | -0.02 | -0.02 | -0.02 | -0.02 | -0.02 | -0.02 | -0.02 |
| | 0.4 | 0.0 | 0.04 | 0.04 | 0.04 | 0.04 | 0.04 | 0.04 | 0.04 | 0.01 | 0.01 | 0.01 | 0.01 | 0.01 | 0.01 | 0.01 |
| | | 2.0 | 0.06 | 0.06 | 0.06 | 0.06 | 0.06 | 0.06 | 0.06 | 0.01 | 0.01 | 0.01 | 0.01 | 0.01 | 0.01 | 0.01 |
| | | 4.0 | 0.10 | 0.10 | 0.10 | 0.10 | 0.10 | 0.10 | 0.10 | 0.01 | 0.01 | 0.01 | 0.01 | 0.01 | 0.01 | 0.01 |
| | 0.6 | 0.0 | 0.05 | 0.05 | 0.05 | 0.05 | 0.05 | 0.05 | 0.05 | 0.00 | 0.01 | 0.00 | 0.00 | 0.00 | 0.01 | 0.00 |
| | | 2.0 | 0.06 | 0.07 | 0.07 | 0.07 | 0.07 | 0.07 | 0.07 | 0.01 | 0.01 | 0.01 | 0.01 | 0.00 | 0.01 | 0.01 |
| | | 4.0 | 0.11 | 0.11 | 0.11 | 0.11 | 0.11 | 0.11 | 0.11 | 0.00 | 0.00 | 0.00 | 0.00 | 0.00 | 0.00 | 0.00 |
| | 0.8 | 0.0 | 0.06 | 0.05 | 0.05 | 0.05 | 0.05 | 0.05 | 0.04 | 0.01 | 0.00 | 0.00 | 0.01 | 0.01 | 0.00 | 0.00 |
| | | 2.0 | 0.08 | 0.07 | 0.06 | 0.06 | 0.06 | 0.07 | 0.06 | 0.01 | 0.01 | 0.01 | 0.01 | 0.01 | 0.01 | 0.01 |
| | | 4.0 | 0.10 | 0.10 | 0.10 | 0.10 | 0.10 | 0.10 | 0.11 | 0.01 | 0.01 | 0.00 | 0.01 | 0.00 | 0.01 | 0.00 |
| 2000 | 0.2 | 0.0 | 0.02 | 0.02 | 0.02 | 0.02 | 0.02 | 0.02 | 0.02 | 0.00 | 0.00 | 0.00 | 0.00 | 0.00 | 0.00 | 0.00 |
| | | 2.0 | 0.03 | 0.03 | 0.03 | 0.03 | 0.03 | 0.03 | 0.03 | 0.00 | 0.00 | 0.00 | 0.00 | 0.00 | 0.00 | 0.00 |
| | | 4.0 | 0.04 | 0.04 | 0.04 | 0.04 | 0.04 | 0.04 | 0.04 | -0.01 | 0.00 | -0.01 | 0.00 | 0.00 | 0.00 | -0.01 |
| | 0.4 | 0.0 | 0.02 | 0.02 | 0.02 | 0.02 | 0.02 | 0.02 | 0.02 | 0.00 | 0.00 | 0.00 | 0.00 | 0.00 | 0.00 | 0.00 |
| | | 2.0 | 0.03 | 0.03 | 0.03 | 0.03 | 0.03 | 0.03 | 0.04 | 0.00 | 0.00 | 0.00 | 0.00 | 0.00 | 0.00 | 0.00 |
| | | 4.0 | 0.05 | 0.05 | 0.05 | 0.05 | 0.05 | 0.05 | 0.05 | 0.00 | 0.00 | 0.00 | 0.00 | 0.00 | 0.00 | 0.00 |
| | 0.6 | 0.0 | 0.02 | 0.02 | 0.02 | 0.02 | 0.02 | 0.02 | 0.02 | 0.00 | 0.00 | 0.00 | 0.00 | 0.00 | 0.00 | 0.00 |
| | | 2.0 | 0.03 | 0.03 | 0.03 | 0.03 | 0.03 | 0.03 | 0.03 | 0.00 | 0.00 | 0.00 | 0.00 | 0.00 | 0.00 | 0.00 |
| | | 4.0 | 0.05 | 0.05 | 0.05 | 0.05 | 0.05 | 0.05 | 0.05 | 0.00 | 0.00 | 0.00 | 0.00 | 0.00 | 0.00 | 0.00 |
| | 0.8 | 0.0 | 0.03 | 0.02 | 0.02 | 0.02 | 0.02 | 0.02 | 0.02 | 0.00 | 0.00 | -0.01 | 0.00 | 0.00 | 0.00 | 0.00 |
| | | 2.0 | 0.04 | 0.04 | 0.04 | 0.04 | 0.04 | 0.04 | 0.04 | 0.00 | 0.01 | 0.00 | 0.00 | 0.00 | 0.00 | 0.00 |
| | | 4.0 | 0.05 | 0.05 | 0.05 | 0.05 | 0.05 | 0.05 | 0.05 | -0.01 | 0.00 | 0.00 | 0.00 | 0.00 | 0.00 | 0.00 |

Table 12: RMSE and Bias of se[$\beta_1$] (LM, LSEM, and SEM). Darker red represents larger RMSE and bias

| N | $\varphi$ | $\tau^2$ | RMSE | | | | | | | Bias | | | | | | |
|---|---|---|---|---|---|---|---|---|---|---|---|---|---|---|---|---|
| | | | LM | LSEM | | | | | SEM | LM | LSEM | | | | | SEM |
| | | | | 50 | 100 | 200 | 300 | 400 | | | 50 | 100 | 200 | 300 | 400 | |
| 500 | 0.2 | 0.0 | 0.002 | 0.002 | 0.002 | 0.002 | 0.002 | 0.003 | 0.001 | 0.001 | 0.000 | -0.001 | -0.001 | -0.001 | -0.002 | 0.000 |
| | | 2.0 | 0.002 | 0.002 | 0.002 | 0.003 | 0.003 | 0.004 | 0.002 | 0.000 | -0.001 | -0.001 | -0.002 | -0.002 | -0.003 | -0.001 |
| | | 4.0 | 0.003 | 0.004 | 0.004 | 0.004 | 0.005 | 0.005 | 0.004 | 0.000 | -0.001 | -0.001 | -0.002 | -0.002 | -0.002 | 0.000 |
| | 0.4 | 0.0 | 0.003 | 0.002 | 0.002 | 0.002 | 0.003 | 0.005 | 0.001 | 0.003 | 0.000 | -0.001 | -0.002 | -0.002 | -0.004 | 0.000 |
| | | 2.0 | 0.002 | 0.002 | 0.003 | 0.003 | 0.004 | 0.005 | 0.002 | 0.001 | -0.001 | -0.002 | -0.002 | -0.002 | -0.003 | 0.000 |
| | | 4.0 | 0.003 | 0.003 | 0.004 | 0.004 | 0.005 | 0.006 | 0.003 | 0.000 | -0.001 | -0.001 | -0.002 | -0.002 | -0.003 | 0.000 |
| | 0.6 | 0.0 | 0.008 | 0.002 | 0.002 | 0.002 | 0.003 | 0.005 | 0.002 | 0.008 | 0.001 | -0.001 | -0.002 | -0.003 | -0.005 | 0.000 |
| | | 2.0 | 0.005 | 0.002 | 0.003 | 0.004 | 0.005 | 0.008 | 0.002 | 0.004 | -0.001 | -0.002 | -0.004 | -0.005 | -0.007 | -0.001 |
| | | 4.0 | 0.004 | 0.004 | 0.004 | 0.005 | 0.005 | 0.007 | 0.003 | 0.002 | -0.001 | -0.002 | -0.003 | -0.003 | -0.005 | 0.000 |
| | 0.8 | 0.0 | 0.024 | 0.005 | 0.002 | 0.002 | 0.004 | 0.006 | 0.002 | 0.023 | 0.004 | 0.000 | -0.001 | -0.003 | -0.006 | 0.000 |
| | | 2.0 | 0.015 | 0.002 | 0.003 | 0.005 | 0.006 | 0.009 | 0.002 | 0.015 | 0.000 | -0.003 | -0.004 | -0.006 | -0.008 | -0.001 |
| | | 4.0 | 0.009 | 0.004 | 0.005 | 0.006 | 0.007 | 0.010 | 0.003 | 0.007 | -0.002 | -0.004 | -0.005 | -0.006 | -0.009 | 0.000 |
| 2000 | 0.2 | 0.0 | 0.001 | 0.000 | 0.000 | 0.000 | 0.001 | 0.001 | 0.000 | 0.000 | 0.000 | 0.000 | 0.000 | 0.000 | 0.000 | 0.000 |
| | | 2.0 | 0.000 | 0.000 | 0.000 | 0.001 | 0.001 | 0.001 | 0.000 | 0.000 | 0.000 | 0.000 | 0.000 | 0.000 | 0.000 | 0.000 |
| | | 4.0 | 0.001 | 0.001 | 0.001 | 0.001 | 0.001 | 0.001 | 0.001 | 0.000 | 0.000 | 0.000 | 0.000 | 0.000 | 0.000 | 0.000 |
| | 0.4 | 0.0 | 0.001 | 0.001 | 0.001 | 0.000 | 0.001 | 0.001 | 0.000 | 0.001 | 0.001 | 0.001 | 0.000 | 0.000 | -0.001 | 0.000 |
| | | 2.0 | 0.001 | 0.001 | 0.001 | 0.001 | 0.001 | 0.001 | 0.001 | 0.001 | 0.000 | 0.000 | 0.000 | -0.001 | -0.001 | 0.000 |
| | | 4.0 | 0.001 | 0.001 | 0.001 | 0.001 | 0.001 | 0.001 | 0.001 | 0.000 | 0.000 | 0.000 | 0.000 | -0.001 | -0.001 | 0.000 |
| | 0.6 | 0.0 | 0.004 | 0.003 | 0.002 | 0.001 | 0.000 | 0.001 | 0.000 | 0.004 | 0.003 | 0.002 | 0.001 | 0.000 | 0.000 | 0.000 |
| | | 2.0 | 0.002 | 0.001 | 0.001 | 0.001 | 0.001 | 0.001 | 0.001 | 0.002 | 0.001 | 0.001 | 0.000 | -0.001 | -0.001 | 0.000 |
| | | 4.0 | 0.001 | 0.001 | 0.001 | 0.001 | 0.001 | 0.001 | 0.001 | 0.001 | 0.000 | 0.000 | -0.001 | -0.001 | -0.001 | 0.000 |
| | 0.8 | 0.0 | 0.012 | 0.007 | 0.005 | 0.002 | 0.001 | 0.000 | 0.000 | 0.012 | 0.007 | 0.005 | 0.002 | 0.001 | 0.000 | 0.000 |
| | | 2.0 | 0.008 | 0.004 | 0.002 | 0.001 | 0.001 | 0.001 | 0.001 | 0.008 | 0.004 | 0.002 | 0.000 | -0.001 | -0.001 | 0.000 |
| | | 4.0 | 0.004 | 0.002 | 0.001 | 0.001 | 0.002 | 0.002 | 0.001 | 0.004 | 0.001 | 0.000 | -0.001 | -0.002 | -0.002 | 0.000 |

Table 13: Bias of the spatial dependence parameter $\varphi$ (LSEM and SEM) and the residual MC values (LM, LSEM, and SEM)

| N | $\varphi$ | $\tau^2$ | Bias of $\varphi$ ||||| | z-value of the residual MC ||||||  |
| | | | LSEM ||||| SLM | LM | LSEM ||||| SEM |
| | | | 50 | 100 | 200 | 300 | 400 | | | 50 | 100 | 200 | 300 | 400 | |
| 500 | 0.2 | 0.0 | 0.12 | 0.32 | 0.50 | 0.54 | 0.54 | -0.02 | 2.73 | 0.31 | -0.20 | 0.08 | -0.02 | -0.22 | 0.10 |
| | | 2.0 | 0.09 | 0.16 | 0.28 | 0.38 | 0.40 | -0.10 | 1.51 | 0.08 | -0.26 | -0.09 | -0.03 | -0.19 | 0.08 |
| | | 4.0 | -0.02 | 0.04 | 0.14 | 0.27 | 0.28 | -0.17 | 0.55 | -0.22 | -0.38 | -0.28 | -0.19 | -0.25 | 0.08 |
| | 0.4 | 0.0 | 0.03 | 0.27 | 0.37 | 0.39 | 0.38 | -0.01 | 6.65 | 1.47 | -0.20 | -0.21 | -0.43 | -0.89 | 0.12 |
| | | 2.0 | 0.04 | 0.21 | 0.35 | 0.38 | 0.36 | -0.18 | 3.36 | 0.30 | -0.56 | -0.26 | -0.34 | -0.55 | 0.06 |
| | | 4.0 | -0.10 | -0.05 | 0.06 | 0.21 | 0.23 | -0.31 | 1.36 | -0.02 | -0.22 | -0.04 | -0.07 | -0.24 | 0.08 |
| | 0.6 | 0.0 | -0.02 | 0.18 | 0.25 | 0.26 | 0.26 | -0.01 | 11.72 | 2.53 | -0.02 | -0.36 | -0.75 | -1.47 | 0.14 |
| | | 2.0 | -0.03 | 0.11 | 0.18 | 0.20 | 0.19 | -0.21 | 6.84 | 0.74 | -0.86 | -1.05 | -1.23 | -1.69 | -0.09 |
| | | 4.0 | 0.03 | 0.06 | 0.14 | 0.20 | 0.19 | -0.41 | 2.93 | -0.16 | -0.89 | -0.70 | -0.67 | -0.88 | 0.01 |
| | 0.8 | 0.0 | 0.03 | 0.08 | 0.12 | 0.13 | 0.13 | 0.00 | 20.77 | 4.86 | 0.65 | -0.22 | -0.88 | -1.77 | 0.17 |
| | | 2.0 | 0.03 | 0.07 | 0.10 | 0.10 | 0.10 | -0.18 | 14.13 | 1.25 | -1.35 | -1.75 | -2.12 | -2.73 | -0.69 |
| | | 4.0 | -0.05 | 0.03 | 0.06 | 0.07 | 0.05 | -0.42 | 6.98 | -0.45 | -1.65 | -1.79 | -1.88 | -2.26 | -0.33 |
| 2000 | 0.2 | 0.0 | 0.17 | 0.20 | 0.22 | 0.29 | 0.41 | 0.00 | 5.89 | 4.33 | 2.90 | 0.74 | -0.72 | -0.76 | 0.04 |
| | | 2.0 | 0.11 | 0.11 | 0.16 | 0.20 | 0.25 | -0.10 | 2.89 | 2.06 | 1.31 | 0.24 | -0.44 | -0.28 | 0.04 |
| | | 4.0 | -0.11 | -0.07 | -0.03 | 0.01 | 0.01 | -0.16 | 1.22 | 0.68 | 0.33 | -0.26 | -0.53 | -0.24 | 0.03 |
| | 0.4 | 0.0 | 0.04 | 0.04 | 0.10 | 0.20 | 0.28 | 0.00 | 13.26 | 9.85 | 6.93 | 2.46 | -0.42 | -0.80 | 0.04 |
| | | 2.0 | 0.01 | 0.06 | 0.10 | 0.18 | 0.26 | -0.17 | 6.96 | 4.74 | 3.00 | 0.40 | -1.16 | -1.30 | -0.04 |
| | | 4.0 | -0.20 | -0.05 | -0.03 | 0.04 | 0.09 | -0.30 | 2.76 | 1.75 | 1.05 | 0.08 | -0.65 | -0.49 | 0.03 |
| | 0.6 | 0.0 | -0.09 | 0.00 | 0.08 | 0.12 | 0.18 | 0.00 | 24.03 | 17.61 | 12.65 | 5.29 | 0.65 | -0.14 | 0.04 |
| | | 2.0 | -0.06 | -0.02 | 0.08 | 0.12 | 0.15 | -0.20 | 13.71 | 9.12 | 5.68 | 0.97 | -1.75 | -2.14 | -0.35 |
| | | 4.0 | -0.16 | -0.03 | 0.03 | 0.07 | 0.13 | -0.41 | 5.80 | 3.30 | 1.64 | -0.48 | -1.66 | -1.75 | -0.10 |
| | 0.8 | 0.0 | -0.08 | 0.02 | 0.06 | 0.07 | 0.09 | 0.00 | 42.35 | 29.49 | 20.70 | 9.17 | 2.38 | 1.05 | 0.03 |
| | | 2.0 | -0.06 | 0.04 | 0.06 | 0.07 | 0.08 | -0.16 | 29.24 | 17.55 | 10.63 | 2.35 | -2.03 | -2.76 | -1.62 |
| | | 4.0 | -0.13 | 0.02 | 0.05 | 0.06 | 0.07 | -0.40 | 14.73 | 7.26 | 3.29 | -1.05 | -3.17 | -3.45 | -0.93 |

Table 14: Computational time for larger samples (seconds). rARPACK package is used for the eigen-decomposition. The bootstrapping is not parallelized.

| N | Eigen-decomposition | | | LSLM | | | | | | LSEM | | |
|---|---|---|---|---|---|---|---|---|---|---|---|---|
| | | | | Estimation | | | Bootstrapping (200 iterations) | | | Estimation | | |
| | $L=50$ | $L=100$ | $L=200$ | $L=50$ | $L=100$ | $L=200$ | $L=50$ | $L=100$ | $L=200$ | $L=50$ | $L=100$ | $L=200$ |
| 5,000 | 1.5 | 2.2 | 5.6 | 0.1 | 0.3 | 1.7 | 10.7 | 27.2 | 139.7 | 0.1 | 0.3 | 1.7 |
| 10,000 | 4.3 | 8.0 | 18.8 | 0.1 | 0.4 | 2.0 | 11.6 | 28.7 | 144.1 | 0.1 | 0.3 | 1.6 |
| 20,000 | 14.3 | 24.3 | 52.0 | 0.2 | 0.6 | 2.4 | 14.0 | 32.4 | 155.1 | 0.1 | 0.5 | 2.3 |
| 40,000 | 46.8 | 72.3 | 148.1 | 0.3 | 0.9 | 3.6 | 19.0 | 40.4 | 176.9 | 0.3 | 0.7 | 2.8 |